\documentclass{article}

\PassOptionsToPackage{numbers,sort&compress}{natbib}
\usepackage[preprint]{neurips_2026}

\usepackage[utf8]{inputenc} 
\usepackage[T1]{fontenc}    
\usepackage{hyperref}       
\usepackage{url}            
\usepackage{booktabs}       
\usepackage{amsfonts}       
\usepackage{nicefrac}       
\usepackage{microtype}      
\usepackage{xcolor}         
\usepackage{graphicx}
\usepackage{amsmath}
\usepackage{algorithm}
\usepackage{algpseudocode}
\usepackage{booktabs}
\usepackage{tabularx}
\usepackage{array}
\usepackage{multirow}
\usepackage{wrapfig}
\usepackage[most]{tcolorbox}
\usepackage{enumitem}
\usepackage{listings}
\usepackage{fancyvrb}
\newcommand{\partitle}[1]{\smallskip \noindent \textbf{#1.}}

\title{What to Remember, What to Reveal: Privacy-Aware Memory for Conversational Agents}

\author{%
  Wenjie Wang$^{*}$ \quad
  Wenhe Si$^{*}$ \quad
  Xinyue Xu \quad
  Yue Xu$^{\dagger}$ \\[0.4em]
  ShanghaiTech University \\[0.2em]
  \texttt{\{wangwj1, siwh2023, xuxy2022, xuyue2022\}@shanghaitech.edu.cn}
}

\begin{document}

\maketitle
\renewcommand{\thefootnote}{\fnsymbol{footnote}}
\footnotetext[1]{Equal contribution.}
\footnotetext[2]{Corresponding author.}
\renewcommand{\thefootnote}{\arabic{footnote}}

\begin{abstract}
Long-term memory enables personalized conversational agents to retain user information across sessions. However,  existing memory architectures primarily optimize for utility but neglect the risks of storing and reusing private attributes such as personally identifiable information (PII) unnecessarily. Dealing with privacy risk in personalized memory is challenging as simply removing sensitive values would undermine the utility of the memory system. Therefore, privacy protection for memory agents must govern the full life-cycle of sensitive values rather than just sanitizing individual records. To fill this research gap, we introduce \textbf{S}anitized \textbf{P}rivacy-\textbf{M}apped M\textbf{em}ory (SP-Mem), a privacy-aware memory architecture that decouples memory utility from exact private-value exposure. 
SP-Mem provides full life-cycle privacy-related design including determining how to identify and separate sensitive information from raw user inputs, how to store sanitized content and exact private values in isolated structures, and how to selectively retrieve values based on the task requirement and user consent. We further introduce a privacy-aware memory benchmark that jointly assesses response quality, privacy behavior, and inference cost. Extensive experiments across multiple LLM-based agents show that SP-Mem achieves stronger personalization while reducing unnecessary privacy exposure. Code and data are available at \url{https://github.com/Jensassss/SP-Mem}.
  

\end{abstract}

\section{Introduction}
\label{sec:introduction}
Long-term memory enables conversational agents to retain user information across sessions~\citep{packer2023memgpt,zhong2024memorybank,park2023generative,wang2023augmenting}, moving beyond isolated prompt-response interactions toward continuous, user-adaptive personalization, which is a fundamental capability in agent design.
However, user memory typically contains both useful non-sensitive preferences and highly sensitive personal information, such as personally identifiable information (PII). 
This mix of information creates a privacy risk: persistent memory may store, retrieve, and reuse private attributes across sessions~\citep{staab2024beyond}, even when irrelevant to the current task~\citep{mextra,ngong2025protecting}. 
The core challenge thus shifts from \textit{enabling memory} to \textit{controlling it}: \textbf{what is remembered, how it is stored, and what is retrieved.}

Existing memory-augmented agent frameworks mainly optimize for utility by persistently extracting user facts, preferences, and summarizing them into a searchable memory architecture~\citep{mem02025,zep2025,li2025memos,secom2025memory,xu2025amem,tan2025rmm}. 
\textbf{This persistent user memory storage introduces a distinct privacy-utility tension: privacy risks arise but simply removing sensitive values would undermine the utility of the memory system.}
On one hand, sensitive information can naturally appear in user-agent interactions~\citep{mextra, ngong2025protecting}. Direct storage in searchable memory is risky~\citep{zeng2024good}, as it can expose private information in unauthorized context, or in situations where only non-sensitive preferences are needed. For example, a user asks for a dinner recommendation based on food preference, which is a task only need non-sensitive preference "taste" but memory systems may retrieve and leak the user's exact home address. On the other hand, exact private values cannot simply be removed or permanently masked, since many personal-assistance tasks require them for task completion, such as form filling or finance- and health-related assistance~\citep{shen2025piibench,pasch2025balancing}. Therefore, privacy protection for memory agents must govern the full life-cycle of sensitive values rather than just sanitize individual prompts.


\begin{figure}
  \centering
  \includegraphics[width=\textwidth,]{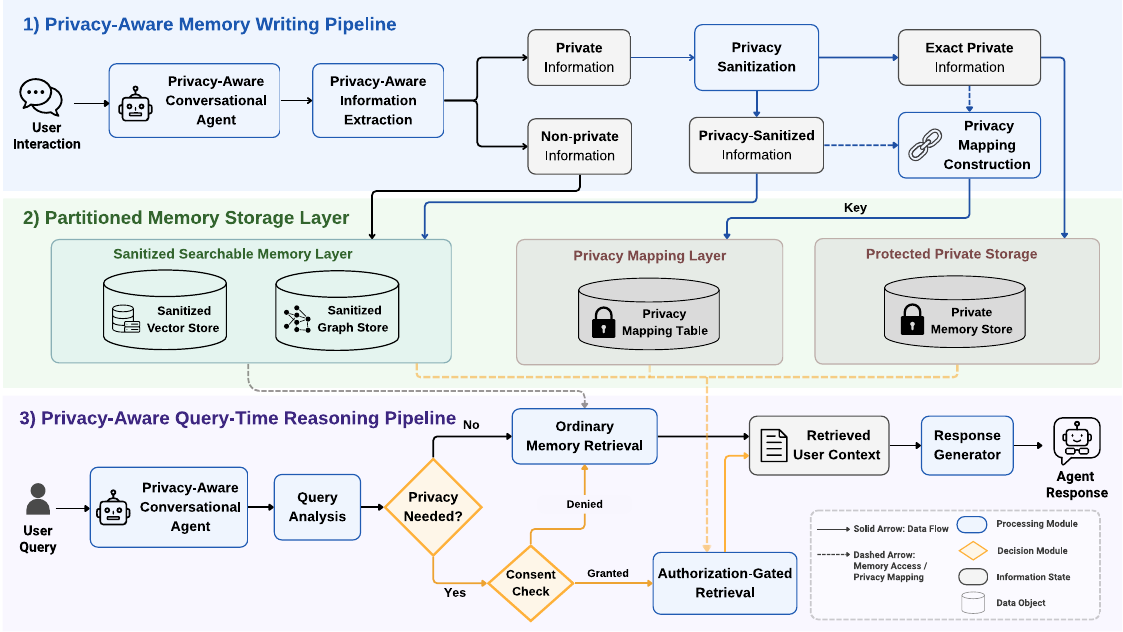}
  \vspace{-1em}
  \caption{Overview of the privacy-aware conversational agent framework. The system 
    contains three stages: privacy-aware memory writing, partitioned memory 
    storage, and query-time privacy-aware reasoning. Private values are sanitized 
    before storage and are only hydrated at inference time when the task requires 
    them and the user grants consent.}
  \label{fig:agent_overview}
\end{figure}

In this work, we propose \textbf{S}anitized \textbf{P}rivacy-\textbf{M}apped M\textbf{em}ory (SP-Mem), a privacy-aware memory architecture that separates searchable sanitized memory from protected private values. As shown in Figure \ref{fig:agent_overview}, it contains three key components. \textit{Privacy-aware memory writing} determines how to identify and separate sensitive information from raw user inputs, \textit{partitioned memory storage} defines how to store sanitized content and exact private values in isolated structures while maintaining a mapping relationship between them, and \textit{privacy-aware query-time reasoning} controls how to selectively retrieve values based on the task requirement and user consent.


To evaluate SP-Mem, we note that existing memory benchmarks primarily emphasize whether the agent can remember and apply user information~\citep{wu2025longmemeval,membench2025,personamem}, but rarely distinguish between appropriate personalization and unnecessary privacy use. An evaluation framework that jointly measures response quality, personalization and privacy-appropriate behavior is needed. Therefore, we introduce a privacy-aware benchmark that includes annotated user profiles, history dialogues, test queries, consent settings, and task-specific information requirements. This yields 2,100 user history dialogues for memory construction and 5,400 queries for evaluation. Experiments across multiple assistant models and memory configurations (including SP-Mem) show that our framework improves personalization while limiting unnecessary privacy exposure. Our contributions can be summarized as follows:
\begin{itemize}
    \item We formulate privacy-aware long-term memory as a core challenge for personalized conversational agents, shifting the focus from remembering user information to controlling how sensitive data is written, stored, retrieved, and used.
    \item We propose SP-Mem, a privacy-aware memory architecture with privacy-preserving memory writing, partitioned memory storage, query-time privacy reasoning, and authorization-gated retrieval,  built on a hybrid graph-vector memory layer.
    \item We introduce the first privacy-aware memory benchmark and evaluation pipeline that jointly assesses response quality, personalization quality, privacy behavior, latency, and token cost. Extensive experiments across multiple LLM-based agents show that SP-Mem achieves stronger personalization while significantly limiting unnecessary privacy exposure.
\end{itemize}

\section{SP-Mem: Sanitized Privacy-Mapped Memory}
\label{sec:privacy_agent}

SP-Mem decouples memory utility from exact private-value exposure via three layers, as illustrated in Figure~\ref{fig:agent_overview}: (1) a privacy-aware memory writing that sanitizes private values during memory extraction; (2) a partitioned memory storage layer that separates sanitized searchable memory from protected mappings to exact private values; and (3) a privacy-aware query reasoning pipeline that selectively retrieves values based on the task requirement and user consent.

\subsection{Privacy-Aware Memory Writing}
\label{sec:privacy_aware_information_extraction}

\begin{figure}
  \centering
  \includegraphics[
   width=\linewidth,
  trim=0cm 0.8cm 0cm 0.1cm,
  clip
  ]{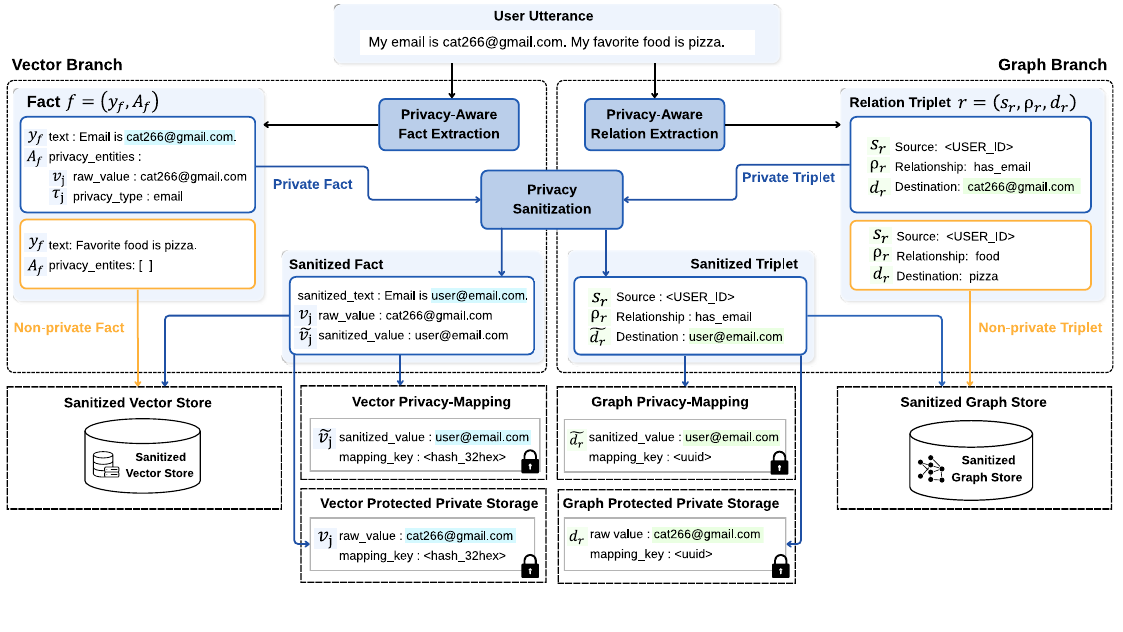}
  \vspace{-1em}
 \caption{Detailed view of SP-Mem. The writing module processes each user input through two parallel branches: a vector branch that extracts fact-style memories and a graph branch that extracts relation triples. Non-private values and sanitized private values are written into the sanitized vector and graph stores. For private values, SP-Mem writes mapping keys into the privacy-mapping layer and stores exact values in the protected private storage.}
  \label{fig:memory_write_detail}
  \vspace{-1em}
\end{figure}

\partitle{Complementary vector and graph memory} As shown in Figure~\ref{fig:memory_write_detail}, SP-Mem uses vector memory and graph memory as complementary retrieval structures. Vector memory stores fact-style information in natural language, enabling semantic retrieval even when the query differs from the stored wording~\citep{lewis2020rag, guu2020realm}. Graph memory stores relation triplets, enabling structured access to user attributes, preferences, and entity-level relations~\citep{edge2025graphrag, han2025graphrag}. By combining the two, SP-Mem can retrieve both broadly relevant facts and relation-specific user information. The following sections detail the privacy extraction and sanitization processes in these two branches.


\partitle{Privacy-aware information extraction} 
Before writing an item into memory, SP-Mem first converts the raw user utterance into two complementary representations: natural-language facts for the vector branch and relation triplets for the graph branch. During this process, privacy-sensitive values are detected and annotated using a rule-based privacy extraction strategy for future sanitization before storage. Let \(\mathcal{T}\) denote a predefined taxonomy of privacy entity types (relegated to Appendix \ref{app:pii_taxonomy}). For the vector branch, the memory writer extracts fact objects $f=(y_f,\mathcal{A}_f)$, where \(y_f\) is a natural-language fact and \(\mathcal{A}_f=\{(v_j,\tau_j)\}_{j=1}^{n}\) is the set of privacy annotations detected in the fact, such as the email address highlighted in light blue on the upper-left of Figure \ref{fig:memory_write_detail}. Each annotation \((v_j,\tau_j)\) consists of a raw private value \(v_j\) and its privacy type \(\tau_j\). Facts with \(\mathcal{A}_f=\emptyset\) are treated as non-private.
For the graph branch, the writer extracts relation triplets
$r=(s_r,\rho_r,d_r)$,
where \(s_r\), \(\rho_r\), and \(d_r\) denote the source entity, relation label, and destination entity, respectively. Relation labels cover privacy-related user attributes, non-private preferences, and general user relations. A triplet is marked as privacy-related if \(\rho_r\) corresponds to privacy-sensitive attributes.


\partitle{Privacy sanitization strategies}
After privacy-aware information extraction, detected private values are transformed into sanitized representations before being written into searchable memory, while non-private facts and triplets remain unchanged. To preserve task-relevant semantics without exposing exact private values, SP-Mem adopts four families of sanitization strategies: name or alias substitution, suffix-preserving masking, numerical bucketing, and LLM-based generalization. Each fine-grained privacy type \(\tau\) is assigned an appropriate strategy according to a type-to-strategy policy \(g\). In the vector branch, each detected private value \(v_j\) in a fact is replaced with its sanitized value \(\tilde{v}_j=\textsc{Sanitize}(v_j,g(\tau_j))\), producing a sanitized fact. For example, the email address in Figure~\ref{fig:memory_write_detail} is sanitized to ``user@email.com''. In the graph branch, triplets with non-private relations remain unchanged, while for privacy-related triplets, the private destination value \(d_r\) is replaced with its sanitized counterpart \(\tilde{d}_r=\textsc{Sanitize}(d_r,g(\tau_r))\). The complete strategy definitions and type-to-strategy mapping are provided in Appendix~\ref{app:privacy_sanitization}.
\vspace{-.5em}
\subsection{Partitioned Storage with Privacy Mapping}
\label{sec:partitioned_storage_with_mapping}
After sanitization, SP-Mem partitions the storage layer into searchable memory, a privacy mapping layer, and a protected private store (the middle part of Figure ~\ref{fig:agent_overview}). The searchable memory consists of vector and graph stores, which contain non-private entries and sanitized representations for general retrieval. Exact private values are stored only in the protected private store. The privacy mapping layer records a mapping key for each sanitized private entry, linking the sanitized representation in searchable memory to its corresponding exact value in the protected store. As a result, exact private values remain outside general retrieval and can be restored only through authorization-gated recovery. A complete specification of this procedure is provided in Algorithm~\ref{alg:privacy_memory_writing} in Appendix~\ref{app:memory_writing}.


\vspace{-.5em}
\subsection{Privacy-Aware Query-Time Reasoning and Authorized Retrieval}
As shown in the bottom part of Figure~\ref{fig:agent_overview}, SP-Mem performs query-time reasoning before memory retrieval. Given a user query, the query analyzer identifies the information required to complete the task and determines whether any required entity corresponds to private information. This step converts the user query into a task-dependent access decision, distinguishing standard retrieval over sanitized searchable memory from authorization-gated retrieval that may restore exact private values.
If no private information is required, SP-Mem retrieves relevant context only from the sanitized searchable memory layer. If private information is required, SP-Mem requests user consent before accessing exact values. With consent, it retrieves relevant sanitized memories and restores only the task-required exact private values through the privacy mapping layer. Without consent, it falls back to sanitized memory only. The response is then generated from the current query and the retrieved context. The prompts used for reasoning and response generation are provided in Appendix \ref{app:inference}.

\vspace{-.5em}
\section{Privacy-Aware Memory Benchmark}
\label{sec:dataset_section}
\begin{figure}[t]
\centering
\includegraphics[
  width=\linewidth,
  trim=0cm 2.6cm 0cm 0.1cm,
  clip
]{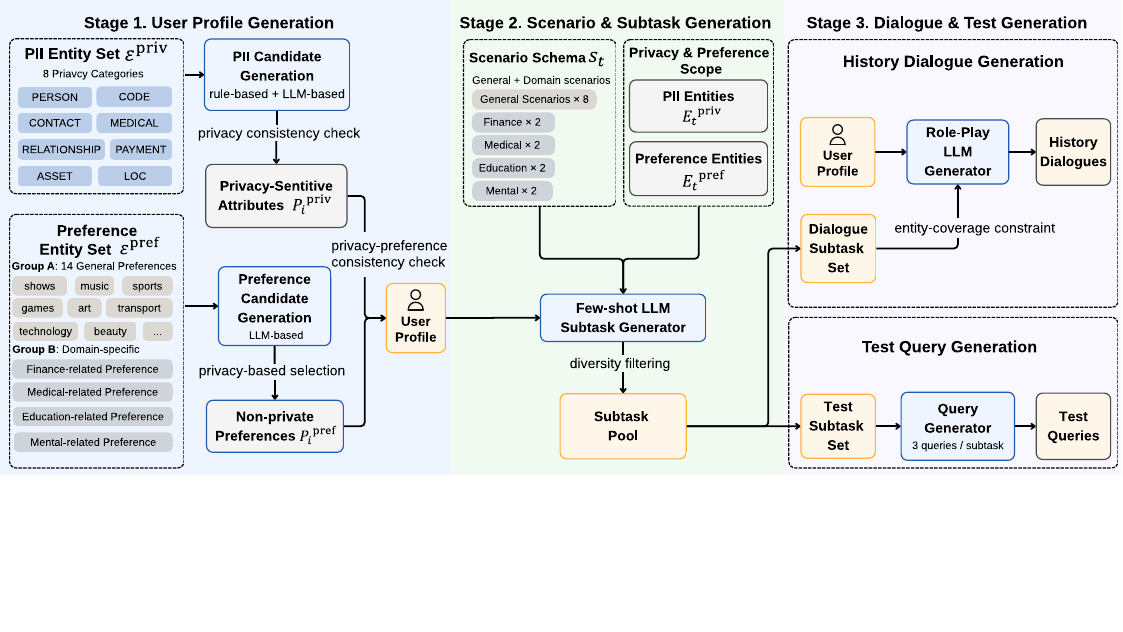}
\vspace{-1em}
\caption{Overview of the benchmark construction pipeline.}
\label{fig:dataset_pipeline}
\vspace{-1em}
\end{figure}

Existing memory benchmarks primarily emphasize whether the agent can remember and apply user information, but rarely distinguish between appropriate personalization and unnecessary privacy use. To evaluate the effectiveness of SP-Mem, we construct the first privacy-aware memory benchmark for assessing response
quality, personalization, and privacy-appropriate behavior in memory-augmented LLM agents.
The benchmark simulates \textbf{long-term personalization memory} by generating user profiles, historical dialogues, and \textbf{scenario-grounded test queries} that involve both privacy-sensitive attributes (the PII Set in Figure \ref{fig:dataset_pipeline}) and non-private preferences (the Preference Entity Set in \ref{fig:dataset_pipeline}). Each test task is annotated with its required information scope, indicating whether the task requires preferences, exact private information, or both, and if privacy is needed, whether user consent is present. This allows us to evaluate whether a memory system can not only improve task completion and personalization, but also decide when private information should be avoided, requested, or restored under user consent.
As illustrated in Figure~\ref{fig:dataset_pipeline}, the benchmark is constructed through a three-stage pipeline. The first stage synthesizes user profiles that contain both privacy-sensitive attributes and non-private personalization preferences. The second stage instantiates scenario-grounded subtasks with explicit privacy and preference scopes. The third stage generates historical dialogues for memory construction and test queries for downstream evaluation.


\vspace{-.5em}
\subsection{User Profile Generation}
The benchmark covers four domains, including finance, medical, education, and mental support, and contains 1,000 synthesized user profiles. Each user profile \(i\) consists of privacy-sensitive attributes \(P^{\mathrm{priv}}_i\) and non-private preferences \(P^{\mathrm{pref}}_i\), which jointly define the user's personal context for dialogue generation and downstream evaluation. For each user, \(P^{\mathrm{priv}}_i\) and \(P^{\mathrm{pref}}_i\) contain concrete values for predefined privacy and preference entity set, denoted by \(\mathcal{E}^{\mathrm{priv}}\) and \(\mathcal{E}^{\mathrm{pref}}\), respectively.

\partitle{Privacy-sensitive attributes} Following prior studies~\citep{shen2025piibench,lukas2023analyzing,sun2024deprompt,quyenanhde_diseases_symptoms}, we define privacy-sensitive attributes as personally identifiable information (PII) and organize the privacy entity set \(\mathcal{E}^{\mathrm{priv}}\) into 8 high-level categories and 37 fine-grained types. The full taxonomy is provided in Appendix~\ref{app:pii_taxonomy}. 
Privacy-sensitive attributes of each user are constructed through a hybrid pipeline. Rule-based generators produce structured fields with explicit formats or consistency constraints, while an LLM enriches open-ended fields conditioned on the structured fields. An LLM-assisted consistency check is further applied to improve completeness and internal consistency.

\partitle{Non-private preferences} Following prior work on preference following and personalized assistants~\citep{prefeval,hicupid}, we construct a preference entity set \(\mathcal{E}^{\mathrm{pref}}\), consisting of 14 general dimensions and 4 domain-specific dimensions. For each dimension, preference candidates are generated with an LLM and refined through semantic deduplication. Each user profile contains a shared set of general preferences and an additional set of preferences specific to its assigned domain. The final non-private preferences are assigned to each user conditioned on \(P^{\mathrm{priv}}_i\), so as to maintain coherence between user background and preferences. A further privacy--preference consistency check is applied to filter conflicting profiles. Additional details on user profile generation are provided in Appendix~\ref{app:dataset}.

\vspace{-1em}
\subsection{Subtask Generation}
Given the synthesized user profiles, this stage defines scenario-grounded subtasks with explicit privacy and preference requirements. We define a scenario schema that covers 8 general scenarios and 8 domain-specific scenarios, with two scenarios for each domain.
For each scenario, a privacy--preference scope is specified by selecting task-relevant privacy entities \(E^{\mathrm{priv}}_t\) and preference entities \(E^{\mathrm{pref}}_t\) from the PII entity set and preference entity set, respectively. These selected entities define the task's information requirements and determine its task mode \(m_t\): \textit{privacy-only} if only privacy entities are selected, \textit{preference-only} if only preference entities are selected, and \textit{mixed} if both are selected. Formally, each subtask requirement is represented as
$
q_t=(s_t,m_t,E^{\mathrm{priv}}_t,E^{\mathrm{pref}}_t),
$ where \(s_t\) denotes the scenario, \(m_t\) is the task mode.

Based on the task requirement \(q_t\) and a user profile, a few-shot LLM generator instantiates concrete subtasks by elaborating the abstract task description into specific user--assistant task settings. To ensure validity and diversity, generated subtasks are filtered if they violate the sampled entity requirements, mismatch the target mode, duplicate existing tasks, or describe implausible user--assistant interactions. The resulting subtask pool serves as the shared source for both history dialogue generation and test query generation in the next stage. After filtering, the pool contains 376 distinct subtasks, including 185 \textit{preference-only} tasks, 51 \textit{privacy-only} tasks, and 140 \textit{mixed} tasks. Full details are provided in Appendix~\ref{app:task_generation}.




\subsection{User History Dialogue and Test Generation}
The third stage generates two types of data from the subtask pool: historical dialogues for memory construction and test queries for downstream evaluation. 
For each scenario, six tasks are reserved for testing, including three \textit{preference-only} tasks and three privacy-related tasks (\textit{mixed} or \textit{privacy-only} tasks), forming the test subtask pool.
The remaining tasks are used as a dialogue subtask pool.
Overall, this stage produces 21,000 historical dialogues for memory construction and 54,000 user-query evaluation instances, derived from 270 unique test query variants.
\vspace{-.5em}
\paragraph{History dialogue generation.}
For each user \(i\), 21 dialogue tasks \(D_i\) are sampled from the dialogue subtask pool under an entity-coverage constraint. Specifically, the selected tasks are required to collectively cover all privacy entities and preference dimensions assigned to the user:
\[
\bigcup_{t\in D_i}\left(E^{\mathrm{priv}}_t \cup E^{\mathrm{pref}}_t\right)
\supseteq
\mathcal{E}^{\mathrm{priv}}_i \cup \mathcal{E}^{\mathrm{pref}}_i,
\quad |D_i|=21.
\]

\vspace{-.5em}
This constraint ensures that the resulting history dialogues contain sufficient memory evidence for both privacy-sensitive attributes and non-private preferences. Given the selected tasks and user profile, multi-turn user--assistant dialogues are generated with role-play LLMs. In each dialogue, the user LLM gradually discloses the task-required entity values, and each user utterance is annotated with the corresponding entity labels. Dialogues continue until the required information is collected and the assistant completes the task. Additional details are provided in Appendix~\ref{app:dialogue_generation}.

\paragraph{Test query generation.}
Using the same task-card schema, 90 test task cards are constructed and each card is rewritten into three natural query variants with an LLM, yielding 270 query variants in total. For each user, test queries are sampled from seven shared general scenarios and two domain-specific scenarios, resulting in 54 test queries per user. The required privacy and preference entities are preserved as ground-truth annotations, enabling controlled evaluation of whether memory systems retrieve and use the appropriate information under each task mode.
\vspace{-.5em}
\section{Experiments}
\label{sec:Experiments} 
The experiments evaluate SP-Mem along the full memory lifecycle. After introducing the experimental configurations, we examine privacy-identification accuracy, response quality, and privacy behavior through comparisons with full-context prompting and existing memory architectures. We further analyze efficiency and ablate the memory branches to assess their complementary roles.
\vspace{-.5em}
\subsection{Configurations}
\label{sec:Experimental_Setup}  

\paragraph{Data and evaluation tasks.}
We uniformly sample 100 users across the four domains for evaluation. For each user, we use 21 user-assistant history dialogues to construct memory and 54 user queries for evaluation, resulting in 2,100 history dialogues and 5,400 evaluation queries in total. Each evaluation query is assigned to one of five evaluation task categories in Table~\ref{tab:evaluation-tasks}, according to its task mode and consent condition.
\begin{table}[!ht]
\centering
\vspace{-.5em}
\caption{Evaluation task category.}
\label{tab:evaluation-tasks}
\small
\setlength{\tabcolsep}{6pt}
\begin{tabular}{p{0.22\linewidth}p{0.72\linewidth}}
\toprule
\textbf{Task} & \textbf{Description }\\
\midrule
Preference-only & Requires preferences only; no private information is needed. \\
Privacy-only-allowed & Requires private information only; user allows privacy usage. \\
Privacy-only-denied & Requires private information only; user denies privacy usage. \\
Mixed-allowed & Requires preferences and private information; user allows privacy usage. \\
Mixed-denied & Requires preferences and private information; user denies privacy usage. \\
\bottomrule
\end{tabular}
\vspace{-1em}
\end{table}

\vspace{-.5em}
\paragraph{Baselines and models.}
We compare against four agent configurations: a \textit{Full-context} baseline, where the response model receives the user's complete interaction history and current query directly in the prompt, and three memory-augmented agents based on Zep~\citep{zep2025}, Mem0~\citep{mem02025}, and MemOS~\citep{li2025memos}. For memory-augmented agents, user histories are first converted into memory, and the current query is answered using retrieved memory evidence. To isolate architectural differences among memory modules, GPT-5.2-Chat~\citep{openai_gpt52_chat} is used as the memory-construction model for all memory-augmented methods. At query time, we evaluate response generation with four backbone LLMs: GPT-5.2-Chat~\citep{openai_gpt52_chat}, Llama-3.1-8B-Instruct~\citep{llama31_8b_instruct}, Qwen3-14B~\citep{qwen3_14b}, and DeepSeek-V3.2~\citep{deepseek_v32}.



\paragraph{Metrics.}
We evaluate both memory-writing accuracy and end-to-end query-time performance. For \textbf{memory writing}, we measure privacy-entity identification using accuracy, precision, and recall to evaluate whether private values are correctly detected before storage. For \textbf{end-to-end query-time performance}, we consider three dimensions: response quality, privacy behavior, and inference cost.
For response quality, we use pairwise Task Completion (\textbf{P-TC}) and pairwise Personalization Quality (\textbf{P-PQ}). P-TC evaluates whether the response satisfies the task requirements and is practically useful, while P-PQ evaluates whether the response meaningfully uses the required preference entities without hallucinating unsupported user information. Both metrics are computed using pairwise LLM-as-a-judge evaluation. We use GPT-4.1~\citep{openai_gpt41} as the judge model; for each query and baseline, the judge compares SP-Mem with the baseline and assigns a win, loss, or tie. We report the win-tie rate \((W+T)/N\), where \(W\), \(T\), and \(N\) denote the number of SP-Mem wins, ties, and total comparisons, respectively.
For privacy behavior, we use Privacy-Appropriate Requesting (\textbf{PAR}) and Unnecessary Privacy Usage (\textbf{UPU}). PAR measures whether the agent detects task-required private information and requests permission before using it; we report accuracy, precision, and recall. UPU is a rule-based binary metric that detects exposure of exact private values that are unnecessary for the task or unauthorized under the consent condition; we report the exposure rate. For inference cost, we report total token usage. More details are provided in Appendix~\ref{app:experimental_setup}.

\subsection{Accuracy of Privacy-Aware Memory Writing}
\begin{wraptable}{r}{0.4\linewidth}
\vspace{-4em}
\centering
\caption{Privacy-entity identification\\performance.}
\vspace{.5em}
\label{tab:privacy_extraction_accuracy}
\small
\setlength{\tabcolsep}{6pt}
\begin{tabular}{lccc}
\toprule
Domain & Acc. & Rec. & Prec. \\
\midrule
Medical & 0.996 & 0.994 & 0.957 \\
Finance & 0.996 & 0.988 & 0.968 \\
Education & 0.997 & 0.992 & 0.975 \\
Mental & 0.996 & 0.994 & 0.959 \\
\midrule
All & 0.996 & 0.992 & 0.965 \\
\bottomrule
\end{tabular}
\vspace{-1.0em}
\end{wraptable}


Accurate privacy identification at the memory-writing stage is critical to SP-Mem because it prevents exact private values from entering searchable memory. 
As shown in Table~\ref{tab:privacy_extraction_accuracy}, SP-Mem achieves consistently strong performance across domains, with an overall accuracy of 0.996, recall of 0.992, and precision of 0.965. These results indicate  that the writing module achieves strong and stable performance across domains, supporting the downstream retrieval and generation stages.

\vspace{-1em}
\subsection{Response Quality}

We evaluate response quality with pairwise P-TC and P-PQ comparisons under two task groups: allowed-access tasks, including Preference-only, Privacy-only-allowed, and Mixed-allowed, and denied-access tasks, including Privacy-only-denied and Mixed-denied. This split evaluates whether SP-Mem maintains response quality both when private information can be used and when exact private values must remain unavailable.

\textbf{SP-Mem improves response quality over memory baselines.}
Table~\ref{table-pairwise-allowed} shows the performance of allowed-access task, where SP-Mem achieves strong P-TC and P-PQ win-tie rates against all memory baselines, with most values above 85\%. 
This indicates that privacy-aware writing preserves enough useful evidence for task completion and personalization through sanitized facts and structured memory links. 
Compared with \textit{Full-context}, SP-Mem is not always stronger on P-TC because \textit{Full-context} gives the response model access to the entire interaction history, which often provides sufficient evidence for task completion. However, this unrestricted context can dilute attention over user preferences and make the model less reliable at following the most relevant personalization signals. As a result, SP-Mem remains especially competitive on P-PQ, suggesting that personalization depends more on retrieving concise and relevant user signals than on exposing the full history.

\begin{table}[!ht]
  \caption{Pairwise response-quality comparison on Preference-only, Privacy-only-allowed, and Mixed-allowed tasks. Values report the win-tie rate of \textbf{SP-Mem} against each baseline.}
  \label{table-pairwise-allowed}
  \centering
  \small
  \newlength{\colwidth}
  \setlength{\colwidth}{1.12cm}
  \begin{tabular*}{\linewidth}{@{\extracolsep{\fill}}
    l
    *{4}{>{\centering\arraybackslash}p{\colwidth} >{\centering\arraybackslash}p{\colwidth}}
    @{}}
    \toprule
    \multirow{2}{*}{\textbf{SP-Mem} vs.}
    & \multicolumn{2}{c}{GPT-5.2-Chat}
    & \multicolumn{2}{c}{Llama-3.1-8B-Instruct}
    & \multicolumn{2}{c}{DeepSeek-V3.2}
    & \multicolumn{2}{c}{Qwen3-14B} \\
    \cmidrule(lr){2-3}
    \cmidrule(lr){4-5}
    \cmidrule(lr){6-7}
    \cmidrule(lr){8-9}
    & P-TC$\uparrow$ & P-PQ$\uparrow$
    & P-TC$\uparrow$ & P-PQ$\uparrow$
    & P-TC$\uparrow$ & P-PQ$\uparrow$
    & P-TC$\uparrow$ & P-PQ$\uparrow$ \\
    \midrule
    Full-context & 65.35\% & 79.13\% & 66.43\% & 85.75\% & 71.34\% & 80.12\% & 45.09\% & 71.70\% \\
    MemOS        & 90.29\% & 94.98\% & 85.88\% & 90.12\% & 94.00\% & 95.65\% & 77.38\% & 90.11\% \\
    Mem0         & 90.87\% & 94.24\% & 87.15\% & 85.87\% & 95.33\% & 95.78\% & 80.63\% & 89.54\% \\
    Zep          & 91.60\% & 94.63\% & 87.02\% & 88.04\% & 95.56\% & 95.61\% & 79.13\% & 89.64\% \\
    \bottomrule
  \end{tabular*}
\end{table}

\textbf{SP-Mem is especially effective when private values are unavailable.}
Table~\ref{table-pairwise-denied} reports results on Privacy-only-denied and Mixed-denied tasks, where exact private values should not be used. SP-Mem achieves win-tie rates above 86\% against all the baselines across available backbones and metrics. This validates SP-Mem's privacy-aware sanitization strategy, showing that sanitized memory can still provide useful task and personalization signals when raw private values are inaccessible.

\begin{table}[!ht]
  \caption{Pairwise response-quality comparison on Privacy-only-denied and Mixed-denied tasks. Values use the win-tie rate of \textbf{SP-Mem} against each baseline.}
  \label{table-pairwise-denied}
  \centering
  \small
  \newlength{\colwidthb}
  \setlength{\colwidthb}{1.12cm}
  \begin{tabular*}{\linewidth}{@{\extracolsep{\fill}}
    l
    *{4}{>{\centering\arraybackslash}p{\colwidthb} >{\centering\arraybackslash}p{\colwidthb}}
    @{}}
    \toprule
    \multirow{2}{*}{\textbf{SP-Mem} vs.}
    & \multicolumn{2}{c}{GPT-5.2-Chat}
    & \multicolumn{2}{c}{Llama-3.1-8B-Instruct}
    & \multicolumn{2}{c}{DeepSeek-V3.2}
    & \multicolumn{2}{c}{Qwen3-14B} \\
    \cmidrule(lr){2-3}
    \cmidrule(lr){4-5}
    \cmidrule(lr){6-7}
    \cmidrule(lr){8-9}
    & P-TC$\uparrow$ & P-PQ$\uparrow$
    & P-TC$\uparrow$ & P-PQ$\uparrow$
    & P-TC$\uparrow$ & P-PQ$\uparrow$
    & P-TC$\uparrow$ & P-PQ$\uparrow$ \\
    \midrule
    Full-context & 91.45\% & 90.86\% & 96.06\% & 96.78\% & 87.14\% & 87.59\% & 86.65\% & 88.88\% \\
    MemOS        & 92.85\% & 94.58\% & 94.23\% & 91.41\% & 95.04\% & 94.11\% & 87.59\% & 91.84\% \\
    Mem0         & 92.74\% & 94.53\% & 93.72\% & 88.00\% & 95.59\% & 93.58\% & 90.52\% & 90.47\% \\
    Zep          & 93.39\% & 94.30\% & 96.17\% & 91.54\% & 95.69\% & 93.58\% & 89.05\% & 90.62\% \\
    \bottomrule
  \end{tabular*}
\vspace{-1em}
\end{table}

\begin{table}[b]
\centering

\begin{minipage}{0.3\linewidth}
\centering
\caption{PAR results across backbone models.}
\label{tab:par-results}
\small
\setlength{\tabcolsep}{4pt}
\begin{tabular}{lcccc}
\toprule
Model & Acc. & Rec. & Prec. & F1 \\
\midrule
GPT-5.2 & 0.89 & 0.84 & 1.00 & 0.91 \\
Llama-3.1 & 0.95 & 0.92 & 1.00 & 0.96 \\
DeepSeek & 0.93 & 0.90 & 1.00 & 0.95 \\
Qwen3 & 0.94 & 0.91 & 1.00 & 0.95 \\
\bottomrule
\end{tabular}
\end{minipage}
\hfill
\begin{minipage}{0.65\linewidth}
\centering
\caption{Relative total token usage normalized by \textit{Full-context}.}
\label{table-efficiency}
\small
\setlength{\tabcolsep}{5pt}
\begin{tabular}{lcccc}
\toprule
Method & GPT-5.2 & Llama-3.1 & DeepSeek & Qwen3 \\
\midrule
Full-context & 1.00 & 1.00 & 1.00 & 1.00 \\
MemOS        & 0.27 & 0.25 & 0.25 & 0.23 \\
Mem0         & 0.25 & 0.23 & 0.24 & 0.21 \\
Zep          & 0.25 & 0.24 & 0.24 & 0.22 \\
\midrule
SP-Mem       & 0.31 & 0.30 & 0.30 & 0.26 \\
\bottomrule
\end{tabular}
\end{minipage}
\end{table}

\subsection{Privacy Behavior}
\begin{figure}[!ht]
  \centering
  \includegraphics[
   width=\linewidth,
   trim=0cm 0.1cm 0cm 0cm,
   clip
  ]{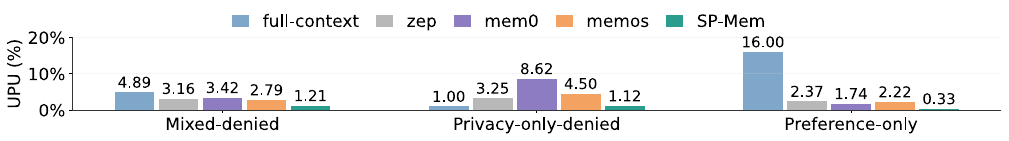}
  \vspace{-1em}
  \caption{UPU across methods in evaluation tasks where exact private values should not appear.}
  \label{fig:UPU_bar}
  \vspace{-1em}
\end{figure}

We evaluate the privacy behavior of SP-Mem using PAR and UPU across different task settings. PAR is evaluated on all five tasks to assess whether the model requests authorization when private information is needed. UPU is evaluated on -denied and Preference-only tasks to examine whether exact private values still appear when authorization is denied or privacy access is unnecessary.
Table~\ref{tab:par-results} reports PAR precision and recall across different backbones. SP-Mem achieves PAR precision of 1.00 across all backbones, with recall between 0.84 and 0.92, suggesting that SP-Mem avoids unnecessary permission requests while still identifying most privacy-requiring tasks.
Figure~\ref{fig:UPU_bar} reports UPU in settings where exact private values should not appear. SP-Mem keeps UPU low in all evaluation tasks, with 1.21\% on Mixed-denied, 1.12\% on Privacy-only-denied, and 0.33\% on Preference-only. 
The contrast is especially clear in Preference-only tasks, where \textit{Full-context} exposes exact private values in 16.00\% of responses, while SP-Mem reduces this rate to 0.33\%. 
Overall, SP-Mem controls privacy at both stages: requesting authorization before private-value access and limiting exact private-value exposure during final response generation.

\subsection{Privacy and Cost Trade-offs}
We compare the relative token usage to assess the cost of SP-Mem and other memory baselines normalized by \textit{Full-context} prompting. Table~\ref{table-efficiency} shows that SP-Mem uses only 26--31\% of the tokens required by Full-context prompting across backbones. Compared with other memory baselines, SP-Mem consumes more tokens, but the added cost supports stronger P-TC and P-PQ performance. This makes the moderate token overhead introduced by SP-Mem acceptable for long-term conversational agents where personalization quality and privacy control are both required.

\subsection{Ablation Study}
To isolate the contribution of each memory branch, we compare the full \textit{vector+graph} memory with two ablated variants: \textit{vector-only}, which only have the vectory branch, and \textit{graph-only}, which only have the graph branch. 
As shown in Table~\ref{tab:ablation-memory}, \textit{vector+graph} achieves win-tie rates above 80\% against both single-branch variants across all backbones and metrics, indicating that the hybrid memory is consistently preferred over either ablation. 
The comparison against both single-branch variants suggests that the two branches are complementary. The gains over \textit{vector-only} are especially strong on P-PQ, indicating that graph memory helps capture structured user-entity relations for preference personalization. The gains over \textit{graph-only} remain consistently high on P-TC, suggesting that vector memory contributes broader semantic context beyond explicit relational triplets.

\begin{table}[!ht]
  \caption{Ablation study of vector and graph memory. Each value reports the win-tie rate of \textbf{\textit{vector+graph}} against a single-branch variant.}
  \label{tab:ablation-memory}
  \centering
  \small
  \setlength{\tabcolsep}{2pt}
  \newlength{\ablationcolwidth}
  \setlength{\ablationcolwidth}{1.15cm}
  \begin{tabular*}{\linewidth}{@{\extracolsep{\fill}}
    l
    *{4}{>{\centering\arraybackslash}p{\ablationcolwidth} >{\centering\arraybackslash}p{\ablationcolwidth}}
    @{}}
    \toprule
    \multirow{2}{*}{\textbf{\textit{vector+graph}} vs.}
    & \multicolumn{2}{c}{GPT-5.2-Chat}
    & \multicolumn{2}{c}{Llama-3.1-8B-Instruct}
    & \multicolumn{2}{c}{DeepSeek-V3.2}
    & \multicolumn{2}{c}{Qwen3-14B} \\
    \cmidrule(lr){2-3}
    \cmidrule(lr){4-5}
    \cmidrule(lr){6-7}
    \cmidrule(lr){8-9}
    & P-TC$\uparrow$ & P-PQ$\uparrow$
    & P-TC$\uparrow$ & P-PQ$\uparrow$
    & P-TC$\uparrow$ & P-PQ$\uparrow$
    & P-TC$\uparrow$ & P-PQ$\uparrow$ \\
    \midrule
    \textit{graph-only}
    & 91.11\% & 88.73\%
    & 85.86\% & 80.37\%
    & 88.55\% & 85.38\%
    & 86.39\% & 82.27\% \\

    \textit{vector-only}
    & 90.77\% & 92.42\%
    & 85.52\% & 84.96\%
    & 88.43\% & 90.65\%
    & 87.23\% & 88.24\% \\
    \bottomrule
  \end{tabular*}
  \vspace{-1em}
\end{table}

\section{Related Work}
\label{sec:relatedworks}
  \vspace{-.5em}
\paragraph{Memory architectures for LLM agents.}

Recent work has increasingly equipped LLM-based agents with external memory for long-term interaction and personalized assistance~\citep{wu2026memory}. 
Early memory streams and long-term memory banks~\citep{park2023generative,zhong2024memorybank} have been extended to structured memory systems, including hybrid vector and graph memory in Mem$0^g$ and Zep~\citep{mem02025,zep2025}, memory-agent management in MemOS~\citep{li2025memos}, and hierarchical memory in MemoryOS~\citep{kang2025memoryos}. 
However, these systems primarily optimize memory utility and downstream task performance, rather than privacy-governed storage, retrieval, and selective disclosure of sensitive user information.

  \vspace{-.5em}

\paragraph{Benchmarks for personalization and privacy.}

Existing benchmarks typically evaluate personalization and privacy separately. Personalization benchmarks test whether LLMs can adapt using user histories or profiles~\citep{lamp,prefeval,hicupid,personamem}. Privacy benchmarks instead focus on query-aware PII protection, contextual privacy awareness, leakage detection, or reconstruction of withheld attributes~\citep{shen2025piibench,privacylens,privlm_bench,propile}. However, long-term memory agents require a more conditional view: private information is not uniformly forbidden, but should be used only when task-relevant and user-authorized. Our benchmark targets this gap by separating non-private preferences from exact private information and evaluating tasks that require preferences, private values, or both under different consent conditions.


  \vspace{-.5em}
\paragraph{Privacy protection methods.}

Existing privacy-protection methods for LLM applications typically detect and transform private entities before inference~\citep{sun2024deprompt,sun2025privprompt}, with prior work studying the privacy--utility trade-off of masking and pseudonymization~\citep{pasch2025balancing}. However, these methods mainly target transient prompts and responses, whereas persistent memory may store, retrieve, and reuse sensitive interactions across future sessions~\citep{mextra}. SP-Mem protects the full memory lifecycle by storing sanitized memories in the searchable layer, keeping exact private values separate, and restoring them only when task-relevant and user-authorized.
Detailed discussion of related work is provided in Appendix~\ref{related_works}.


  \vspace{-.5em}
\section{Conclusion}
\label{sec:conclusion}

We introduced \textbf{S}anitized \textbf{P}rivacy-\textbf{M}apped M\textbf{em}ory (SP-Mem), a privacy-aware memory architecture for long-term conversational agents. SP-Mem separates searchable sanitized memory from protected exact private values, and restores private information only when it is both task-required and user-authorized. We also built a privacy-aware personalization benchmark across four domains to jointly evaluate response utility, personalization quality, privacy behavior, and inference cost. Experiments across multiple LLM backbones show that SP-Mem maintains strong personalized assistance while reducing unnecessary exposure of exact private values. While our benchmark provides a controlled testbed for privacy-aware personalization, future work should examine SP-Mem in real-world interactions and support finer-grained user-specific privacy preferences.

\newpage
\bibliographystyle{unsrtnat}
\bibliography{references}

\newpage
\appendix

\section{Limitations}
\label{sec:limitations}

SP-Mem is evaluated in a controlled setting with synthetic user profiles, dialogues, and task queries. This design enables systematic evaluation across privacy entities, user preferences, task modes, and authorization conditions. However, real user interactions can be more ambiguous, incomplete, and temporally evolving than our benchmark setting. Future work should evaluate privacy-aware memory systems in more naturalistic or user-validated scenarios, while minimizing the collection and exposure of sensitive information during evaluation.

Our privacy analysis focuses on exact private-value exposure and authorization-aware memory use. Specifically, UPU measures whether an agent response contains exact private values when such information is unnecessary or not authorized. This metric captures a central failure mode of memory-augmented personalized agents, but it does not cover all privacy risks. For example, generalized attributes may still support re-identification, multi-turn interactions may enable inference attacks, and adversaries may attempt to recover sanitized values. These limitations suggest that privacy-aware memory systems can improve user control and reduce unnecessary exposure, but such systems should be evaluated under stronger threat models and more realistic settings in future work.

\section{Detailed Related Works}
\label{related_works}

\paragraph{Memory architectures for LLM agents.}
Recent memory-augmented LLM agents mainly study how to store, retrieve, and update past interactions to support long-term continuity, personalization, planning, and context management. Generative Agents~\citep{park2023generative} maintain a natural-language memory stream for reflection and planning, while MemoryBank~\citep{zhong2024memorybank} stores dialogue history, event summaries, and user portraits for long-term personalization. MemGPT~\citep{packer2023memgpt} frames memory as context management and uses an OS-inspired hierarchy to move information between in-context memory and external archival storage. More recent systems, such as MemoryOS~\citep{kang2025memoryos} and Mem0~\citep{mem02025}, package memory extraction, updating, retrieval, and response generation into general-purpose memory infrastructure for LLM agents.

These systems also differ in memory representation. Vector-based memory stores conversations or extracted facts as dense embeddings for semantic retrieval, as in MemoryBank~\citep{zhong2024memorybank} and A-MEM~\citep{xu2025amem}; A-MEM further augments memory notes with attributes and dynamic links. Graph-based memory represents information as entities, relations, and temporal structures: Zep~\citep{zep2025} builds a temporal knowledge graph for evolving agent memory, and AriGraph~\citep{arigraph2024} integrates semantic and episodic memories into a graph for structured retrieval and decision making. Hybrid systems such as Mem0~\citep{mem02025} combine vector retrieval with graph-enhanced representations, suggesting that semantic recall and relational structure provide complementary benefits.

However, these architectures primarily optimize what agents can remember and reuse across long-term interactions. In contrast, our work focuses on privacy-aware memory lifecycle management: determining what should be written into searchable memory, what should remain protected outside it, and when exact private values can be restored under task necessity and user authorization.

\paragraph{Datasets and benchmarks for personalization and privacy.}
As LLMs are increasingly deployed as personalized assistants, recent benchmarks evaluate whether models can use user-specific profiles, histories, and preferences across tasks and conversations. LaMP~\citep{lamp} studies profile-conditioned personalized language modeling across classification and generation tasks, while PrefEval~\citep{prefeval} focuses on explicit and implicit preference following in multi-session conversations. HiCUPID~\citep{hicupid} and PersonaMem~\citep{personamem} further move personalization toward assistant-style interactions, evaluating whether models can ground responses in user backgrounds, conversational context, and evolving user profiles.

In parallel, privacy-oriented benchmarks evaluate whether LLMs leak sensitive information or follow appropriate privacy norms. PII-Bench~\citep{shen2025piibench} studies query-aware PII protection by distinguishing query-relevant from query-unrelated PII. PrivacyLens~\citep{privacylens} evaluates contextual privacy awareness in privacy-sensitive agent scenarios. PrivLM-Bench~\citep{privlm_bench} measures privacy leakage under attacks such as membership inference, training data extraction, and embedding inversion, while ProPILE~\citep{propile} probes whether models can reconstruct withheld personal attributes from related PII.

These benchmarks provide important tools for evaluating either personalization quality or privacy risk. However, they typically treat personalization and privacy as separate objectives. Our work targets their intersection: privacy-aware personalization with long-term memory. We construct user profiles containing both non-private preferences and privacy-sensitive attributes, and design tasks that require preference adaptation, private information usage, or both. This setting evaluates whether agents can personalize responses while avoiding unnecessary access to or disclosure of exact private information.

\paragraph{Privacy protection methods.}
Existing privacy-protection methods for LLM applications mainly operate at the prompt/input level. These methods detect private entities in user inputs and transform them before model inference. DePrompt~\citep{sun2024deprompt} and PrivPrompt~\citep{sun2025privprompt} identify PII in LLM prompts and use generative sanitization to preserve task-relevant semantics while weakening the link between identifiers and sensitive attributes. Complementary studies analyze the privacy-utility trade-off of different anonymization strategies. For example, prior work compares masking, contextual masking, and pseudonymization in personal writing tasks, showing that privacy protection must preserve enough context for useful LLM-generated responses~\citep{pasch2025balancing}. 

Beyond prompt-level protection, recent work shows that long-term agent memory can itself become a privacy attack surface. MEXTRA~\citep{mextra} demonstrates that private user-agent interactions stored in memory can be extracted through black-box attacks. These results suggest that privacy protection cannot be treated only as a preprocessing step before inference. In contrast to prompt-level sanitization and post-hoc privacy risk analysis, our work targets the full memory lifecycle of personalized agents: SP-Mem writes sanitized memories into the default searchable layer, keeps exact private values in protected storage, and restores them only under task relevance and user authorization.

\section{Benchmark Details}
\label{app:dataset_details}
In this section, we provide additional details on the construction of the benchmark.

\subsection{User Profile Generation}
\label{app:dataset}
We present the definition of personally identifiable information (PII) and the full PII taxonomy in Appendix~\ref{app:PII_details} and Appendix~\ref{app:pii_taxonomy}. 
Next, we provide the detailed preference taxonomy in Appendix~\ref{app:preference_schema}, including both general preferences and domain-specific preferences. 
The LLM-based open-ended privacy attributes were generated using Gemma-3-12B-IT. 
Figure~\ref{fig:preference_selection_prompt} shows the prompt used to select preference values conditioned on each user's privacy profile. 
Figure~\ref{fig:privacy_consistency_prompt} shows the prompt used to check internal consistency within privacy profiles, and Figure~\ref{fig:privacy_preference_consistency_prompt} shows the prompt used to check consistency between privacy attributes and assigned preferences.

\subsubsection{PII Definition}
\label{app:PII_details}

Following prior work on PII identification, privacy leakage, and anonymization in language-model systems~\citep{shen2025piibench,lukas2023analyzing,sun2024deprompt}, we define personally identifiable information (PII) as user-specific information that can directly identify an individual, increase identifiability when combined with other attributes, or reveal sensitive personal states. 
We categorize PII into three categories:

\begin{itemize}
    \item \textbf{Direct identifiers.} 
    Direct identifiers are PII attributes that can be used to uniquely identify or contact an individual on their own. Examples of direct identifers include names, national ID numbers, passport numbers, phone numbers, and more.

    \item \textbf{Quasi identifiers.} 
     Quasi identifiers are PII attributes that may not uniquely identify an individual alone but can increase identifiability when combined with other information. Examples include age, gender, occupation, and more.

    \item \textbf{Confidential attributes.} 
    Confidential attributes refer to sensitive personal information whose disclosure may significantly affect an individual's privacy, dignity, or security. Examples include medical conditions, family status, and other sensitive personal records.
\end{itemize}

\subsubsection{PII Taxonomy}
\label{app:pii_taxonomy}

Following prior PII taxonomies and privacy benchmarks~\citep{sun2025privprompt,shen2025piibench}, we adapt the entity space to our privacy-aware memory setting. 
We retain entity types that are commonly used in prior work and further organize them according to their role in conversational memory, including identity, contactability, medical status, relationship context, payment records, financial assets, and location information. 
Our taxonomy comprises eight high-level categories and 37 fine-grained entity types, as summarized in Table~\ref{tab:pii_types}.

\textbf{PERSON:} Refers to basic personal attributes that describe an individual, including name, age, gender, nationality, occupation, and education.

\textbf{CODE:} Encompasses official identifiers and government-issued codes, including ID numbers and passport numbers.

\textbf{CONTACT:} Covers information that can be used to contact an individual, including phone numbers and email addresses.

\textbf{MEDICAL:} Contains health-related information, including diagnosis, symptoms, treatments, allergies, medical examinations, and surgical history.

\textbf{RELATIONSHIP:} Represents family and relationship-related attributes, including marital status and number of children.

\textbf{PAYMENT:} Includes payment and transactional records, such as transactions, bank accounts, credit cards, tax IDs, tax payments, and insurance records.

\textbf{ASSET:} Captures financial status and asset-related attributes, including monthly income, monthly expenses, account balance, loan amount, credit limit, debt ratio, investment return, ROI, finance status level, net worth, and credit score.

\textbf{LOC:} Covers location-related information, including home address and work address.

\begin{table}[t]
\centering
\caption{PII taxonomy and generation rules used in our dataset. The taxonomy covers 8 high-level categories and 37 fine-grained entity types. Rule-based fields follow predefined format constraints, while medical fields are seeded from an external medical dataset~\citep{quyenanhde_diseases_symptoms} and LLM-based fields are generated as contextually appropriate textual content.}
\label{tab:pii_types}
\label{tab:pii_types}
\label{tab:pii_types}
\begin{tabularx}{\textwidth}{p{0.18\textwidth} p{0.2\textwidth} p{0.17\textwidth} X}
\toprule
\textbf{PII Category} & \textbf{Entity Type} & \textbf{Generation Source} & \textbf{Format Constraints} \\
\midrule
\small
\multirow{6}{*}{PERSON}
& name & Rule-based & [First Name] [Last Name] \\
& age & Rule-based & [18--90] \\
& gender & Rule-based & Male/Female \\
& nationality & Rule-based & Predefined list \\
& occupation & Rule-based & Predefined list \\
& education & Rule-based & Predefined list \\

\midrule
\multirow{2}{*}{CODE}
& id number & Rule-based & [A-Z]\{2\}-ID-[0-9]\{8\} \\
& passport number & Rule-based & [A-Z]\{2\}-P[0-9]\{9\} \\

\midrule
\multirow{2}{*}{CONTACT}
& phone number & Rule-based & +[0-9]\{1,3\}-[0-9]\{8,11\} \\
& email & Rule-based & [user]@[domain] \\

\midrule
\multirow{6}{*}{MEDICAL}
& diagnosis & Dataset-based & -- \\
& symptoms & Dataset-based & -- \\
& treatments & Dataset-based & -- \\
& allergies & Dataset-based & -- \\
& exams & Dataset-based & -- \\
& surgical history & Dataset-based & -- \\

\midrule
\multirow{2}{*}{RELATIONSHIP}
& marriage & Rule-based & Single/Married/Divorced/Widowed \\
& children count & Rule-based & [0--4] \\

\midrule
\multirow{6}{*}{PAYMENT}
& transactions & LLM-based & -- \\
& tax payment & Rule-based & [Currency][0-9]+.[0-9]\{2\} \\
& bank account & Rule-based & [0-9]\{12\} \\
& credit card & Rule-based & [0-9]\{16\} \\
& tax ID & Rule-based & [0-9]\{10\} \\
& insurance record & LLM-based & -- \\

\midrule
\multirow{11}{*}{ASSET}
& monthly income & Rule-based & [Currency][0-9]+.[0-9]\{2\} \\
& monthly expenses & Rule-based & [Currency][0-9]+.[0-9]\{2\} \\
& account balance & Rule-based & [Currency][0-9]+.[0-9]\{2\} \\
& loan amount & Rule-based & [Currency][0-9]+.[0-9]\{2\} \\
& credit limit & Rule-based & [Currency][0-9]+.[0-9]\{2\} \\
& debt ratio & Rule-based & [0--0.9999] \\
& investment return & Rule-based & [Currency][0-9]+.[0-9]\{2\} \\
& ROI & Rule-based & [-0.15--0.25] \\
& finance status level & Rule-based & Very Low/Low/Medium/High/Very High \\
& net worth & Rule-based & [Currency][0-9]+.[0-9]\{2\} \\
& credit score & Rule-based & [300--850] \\

\midrule
\multirow{2}{*}{LOC}
& home address & LLM-based & -- \\
& work address & LLM-based & -- \\

\bottomrule
\end{tabularx}
\end{table}

\subsubsection{Preference Taxonomy}
\label{app:preference_schema}

Following PrefEval~\citep{prefeval} and HiCUPID~\citep{hicupid}, we construct a non-private preference taxonomy for controlled user-profile generation. 
We align scene-based preference categories with fine-grained persona dimensions, while filtering out sensitive information such as demographic, identity, financial personal information. 
The resulting taxonomy contains 14 general preference dimensions and 16 domain-specific preference dimensions across four domains.

\paragraph{General preference dimensions.}
The general preference dimensions capture everyday interests, tastes, and lifestyle preferences:
shows, music, books, games, art, sports, fitness, food, beauty, clothing, technology, transport, travel, and pets.

\paragraph{Domain-specific preference dimensions.}
The domain-specific preference dimensions capture personalization signals required by specialized scenarios:
\begin{itemize}
    \item \textbf{Finance:} risk tolerance, financial news source preference, financial market sector preference, and sustainability preference.
    \item \textbf{Medical:} medical decision role, medical tone preference, medical risk attitude, and medical follow-up reminder preference.
    \item \textbf{Education:} learning style preference, learning resource preference, learning schedule preference, and learning feedback preference.
    \item \textbf{Mental support:} coping strategy preference, mental health topic preference, mental health support response preference, and stress response tendency.
\end{itemize}

\subsection{Subtask Generation Details}
\label{app:task_generation}

This section provides supplementary details for the entity-controlled subtask generation process described in Section~\ref{app:subtask_generation}. 
We first present the full scenario taxonomy in Section~\ref{app:scenario_taxonomy}. 
Figure~\ref{fig:task_card_generation_prompt} shows the prompt template used for constrained task-card generation, where the scenario, task mode, and sampled entity annotations are fixed before LLM generation. 
After generation, we manually inspect and refine the generated cards to ensure that they are realistic, mode-consistent, and aligned with the assigned entity annotations.

\subsubsection{Scenario Taxonomy}
\label{app:scenario_taxonomy}

We define scenarios to provide concrete contexts for task-card generation. 
Table~\ref{tab:scenario_taxonomy} summarizes 16 scenarios, including 8 shared general scenarios and 8 domain-specific scenarios across finance, medical, education, and mental-support domains. 
General scenarios can be instantiated with any privacy entity in \(\mathcal{E}^{\mathrm{priv}}\) and the general preference dimensions. 
Domain-specific scenarios can also use any privacy entity in \(\mathcal{E}^{\mathrm{priv}}\), but their preference dimensions are restricted to the corresponding domain. 
For example, finance scenarios use finance-specific preferences and are assigned only to finance-domain users; the same rule applies to medical, education, and mental-support scenarios. 
We additionally define one auxiliary profile-completion scenario for dialogue generation only.

\begin{table}[t]
\centering
\caption{Scenario taxonomy used for task-card generation.}
\label{tab:scenario_taxonomy}
\begin{tabular}{p{0.22\linewidth} p{0.30\linewidth} p{0.38\linewidth}}
\toprule
\textbf{Scenario group} & \textbf{Scenario} & \textbf{Description} \\
\midrule
General 
& Writing and communication 
& Drafting, polishing, or revising messages, emails, invitations, and customer-service communication. \\

& Planning and scheduling 
& Creating plans, schedules, reminders, to-do lists, and time arrangements for everyday activities. \\

& Local life and services 
& Handling local service needs such as appointments, deliveries, refunds, repairs, and daily-life logistics. \\

& Account, forms, and verification 
& Organizing information for accounts, applications, verification forms, document lists, and administrative workflows. \\

& Travel and booking help 
& Supporting travel planning, booking, itinerary comparison, and transportation arrangements. \\

& Wellness and lifestyle 
& Providing lifestyle support for sleep, exercise, diet, habits, routines, and everyday wellness planning. \\

& Family and relationship logistics 
& Assisting with family coordination, relationship communication, and interpersonal planning. \\

& Profile completion 
& Completing user profile information for dialogue generation. \\

\midrule
Finance 
& Personal finance operations 
& Supporting budgeting, bill organization, repayment planning, and everyday financial management. \\

& Investing and wealth 
& Supporting investment planning, asset allocation, and wealth-management decisions. \\

\midrule
Medical 
& Urgent care and navigation 
& Helping with healthcare navigation, hospital selection, appointment procedures, visit planning, and medication access. \\

& Guidance and understanding 
& Explaining health information, symptoms, examination results, reports, and treatment-related questions. \\

\midrule
Education 
& Education planning and academic logistics 
& Supporting study plans, review schedules, course selection, applications, enrollment, and academic communication. \\

& Learning support and academic guidance 
& Explaining concepts, providing learning feedback, preparing for exams, and giving personalized study suggestions. \\

\midrule
Mental support 
& Internal emotional struggles 
& Supporting users with stress, anxiety, emotional overwhelm, self-doubt, and decision-related distress. \\

& Psychological topics and life context 
& Exploring self-understanding, relationship patterns, identity-related questions, and the psychological impact of daily life. \\

\bottomrule
\end{tabular}
\end{table}

\subsubsection{Entity-Controlled Task Generation}
\label{app:subtask_generation}
For each scenario and task mode, we first sample the required privacy entity set \(E^{\mathrm{priv}}_t\) and preference entity set \(E^{\mathrm{pref}}_t\) from the corresponding allowlists. 
The sampled entity sets are then fixed in the prompt: the LLM is instructed to generate only the natural-language task description and user intent while copying the sampled entity lists exactly. 
This design ensures that the ground-truth entity annotations are controlled by the dataset construction pipeline rather than freely chosen by the LLM.

The prompt template is shown in Figure~\ref{fig:task_card_generation_prompt}. 
It enforces mode-specific solvability constraints: for \textit{privacy-only} tasks, the task must be solvable using only the selected privacy entities and must not require preferences; for \textit{preference-only} tasks, the task must be solvable using only the selected preference entities and must not require private information; and for \textit{mixed} tasks, the task must require both selected privacy entities and selected preference entities. 
The prompt also requires every selected entity to be necessary for task completion, so that removing any selected entity would make the task incomplete or underspecified.

\subsection{Dialogue Generation}
\label{app:dialogue_generation}

We generate history dialogues to seed memory evidence for each user through a two-level entity coverage process. 
First, we assign dialogue tasks so that the selected task set jointly covers the target privacy and preference entities for each user; this user-level assignment procedure is described in Section~\ref{app:dialogue_task_assignment}. 
Second, for each assigned task, we generate a multi-turn role-play dialogue using two LLMs, with Llama-3.1-8B-Instruct acting as the user and Gemma-3-12B-IT acting as the assistant. 
This task-level coverage procedure ensures that all task-required entities are disclosed in the user's turns, as described in Section~\ref{app:dialogue_coverage}. Token statistics of the generated dialogue histories are reported in Table~\ref{tab:dialogue-token-stats}.

\subsubsection{Dialogue Task Assignment}
\label{app:dialogue_task_assignment}

For each user \(i\), we assign dialogue tasks before generating history dialogues. 
Each task card \(t\) contains required privacy entities \(E^{\mathrm{priv}}_t\) and preference entities \(E^{\mathrm{pref}}_t\). 
We define the target entity set for user \(i\) as
\[
G_i = \mathcal{E}^{\mathrm{priv}}_i \cup \mathcal{E}^{\mathrm{pref}}_i .
\]
The assignment consists of two parts. 
First, we construct a coverage set \(D_i^{\mathrm{cov}}\) of 12 tasks. 
We maintain an uncovered entity set and iteratively assign tasks that cover the largest number of remaining entities, so that the 12 selected tasks collectively cover all entities in \(G_i\). 
Second, we randomly sample 9 additional unused tasks \(D_i^{\mathrm{rand}}\) from the dialogue subtask pool to improve scenario diversity. 
The final dialogue task set is
\[
D_i = D_i^{\mathrm{cov}} \cup D_i^{\mathrm{rand}},
\quad |D_i^{\mathrm{cov}}|=12,\quad |D_i^{\mathrm{rand}}|=9,\quad |D_i|=21.
\]
Since \(D_i^{\mathrm{cov}}\subseteq D_i\), this also implies the coverage constraint stated in the main text.
Algorithm~\ref{alg:task_entity_coverage} summarizes this assignment procedure.

\begin{algorithm}
\caption{Dialogue task assignment with entity coverage}
\label{alg:task_entity_coverage}
\begin{algorithmic}[1]
\Require User \(i\), dialogue task-card pool \(\mathcal{T}\), target entity set \(G_i\), coverage size \(K_{\mathrm{cov}}=12\), total size \(K=21\)
\Ensure Assigned dialogue task set \(D_i\)
\State \(D_i^{\mathrm{cov}} \gets \emptyset\)
\State \(R \gets G_i\) \Comment{uncovered entities}
\While{\(R \neq \emptyset\)}
    \State \(t^\star \gets \arg\max_{t \in \mathcal{T}} |\mathrm{Ent}(t) \cap R|\)
    \State \(D_i^{\mathrm{cov}} \gets D_i^{\mathrm{cov}} \cup \{t^\star\}\)
    \State \(R \gets R \setminus \mathrm{Ent}(t^\star)\)
    \State \(\mathcal{T} \gets \mathcal{T} \setminus \{t^\star\}\)
\EndWhile
\While{\(|D_i^{\mathrm{cov}}| < K_{\mathrm{cov}}\)}
    \State Randomly sample an unused task \(t\) from \(\mathcal{T}\)
    \State \(D_i^{\mathrm{cov}} \gets D_i^{\mathrm{cov}} \cup \{t\}\)
    \State \(\mathcal{T} \gets \mathcal{T} \setminus \{t\}\)
\EndWhile
\State \(D_i^{\mathrm{rand}} \gets \emptyset\)
\While{\(|D_i^{\mathrm{rand}}| < K - K_{\mathrm{cov}}\)}
    \State Randomly sample an unused task \(t\) from \(\mathcal{T}\)
    \State \(D_i^{\mathrm{rand}} \gets D_i^{\mathrm{rand}} \cup \{t\}\)
    \State \(\mathcal{T} \gets \mathcal{T} \setminus \{t\}\)
\EndWhile
\State \(D_i \gets D_i^{\mathrm{cov}} \cup D_i^{\mathrm{rand}}\)
\State \Return \(D_i\)
\end{algorithmic}
\end{algorithm}

\subsubsection{Dialogue Entity Coverage Procedure}
\label{app:dialogue_coverage}

After assigning dialogue tasks at the user level, we generate one multi-turn dialogue for each assigned task. 
While Algorithm~\ref{alg:task_entity_coverage} ensures that the selected task set covers the target entities for a user, Algorithm~\ref{alg:dialogue_coverage} ensures that the required entities of each individual task are disclosed in the generated dialogue. 
The first user turn states only the task intent and does not reveal any concrete profile value. 
In later turns, the user is allowed to disclose one to three new entities from the remaining set. 
A label is accepted only if it belongs to the task's required entity set and has not been disclosed before. 
The dialogue terminates only after all required entities have been covered.

\begin{algorithm}
\caption{Task-level entity coverage for history dialogue generation}
\label{alg:dialogue_coverage}
\begin{algorithmic}[1]
\Require User \(i\), task \(t\), user profile \(U_i\), required privacy entities \(E^{\mathrm{priv}}_t\), required preference entities \(E^{\mathrm{pref}}_t\)
\Ensure Multi-turn history dialogue \(H_{i,t}\)
\State \(E_t \gets E^{\mathrm{priv}}_t \cup E^{\mathrm{pref}}_t\)
\State \(C \gets \emptyset\) \Comment{covered entities}
\State Generate initial user turn \(u_0\) from the task intent without revealing concrete values
\State \(H_{i,t} \gets [u_0]\)
\While{\(C \neq E_t\)}
    \State \(R \gets E_t \setminus C\) \Comment{remaining entities}
    \State Generate assistant turn \(a_k\) that acknowledges prior information and prompts continuation
    \State Generate next user turn \(u_k\) that discloses 1--3 entities from \(R\)
    \State Extract labels \(L_k\) from \(u_k\)
    \State \(L_k \gets \{e \in L_k \mid e \in R\}\) \Comment{accept only valid new labels}
    \State \(C \gets C \cup L_k\)
    \State Append \(a_k\) and \(u_k\) to \(H_{i,t}\)
\EndWhile
\State Generate final assistant response \(a_{\mathrm{final}}\) conditioned on all disclosed entities
\State Append \(a_{\mathrm{final}}\) to \(H_{i,t}\)
\State \Return \(H_{i,t}\)
\end{algorithmic}
\end{algorithm}

\begin{table}
  \caption{Statistics of user history dialogues across domains. 
Token counts are computed over message content using the \texttt{cl100k\_base} tokenizer, excluding chat-template overhead. 
Avg. tokens/user is the average total token count of all history dialogues for each user.}
  \label{tab:dialogue-token-stats}
  \centering
  \setlength{\tabcolsep}{5pt}
  \begin{tabular*}{\linewidth}{@{\extracolsep{\fill}} lcccc @{}}
    \toprule
    Domain & Users & Dialogues & Avg. tokens/dialogue & Avg. tokens/user \\
    \midrule
    Education & 250 & 5250 & 412.74 & 8667.48 \\
    Finance   & 250 & 5250 & 406.14 & 8528.92 \\
    Medical   & 250 & 5250 & 408.19 & 8571.95 \\
    Mental    & 250 & 5250 & 397.83 & 8354.45 \\
    \midrule
    Overall   & 1000 & 21000 & 406.22 & 8530.70 \\
    \bottomrule
  \end{tabular*}
\end{table}

\section{SP-Mem Details}
\label{app:agent_implementation}

\subsection{Privacy-Aware Memory Writing and Partitioned Storage}
\label{app:memory_writing}

Algorithm~\ref{alg:privacy_memory_writing} summarizes the writing-time procedure of SP-Mem. 
Given a user input \(x_t\), SP-Mem converts the input into two complementary representations: fact objects for the vector branch and relation triplets for the graph branch. 
It then sanitizes detected private values before writing them into searchable memory, while storing mapping keys in the privacy mapping layer and exact private values in the protected private store.

Here, \(M_{\mathrm{vector}}\) and \(M_{\mathrm{graph}}\) denote the sanitized searchable vector and graph stores, respectively. 
\(M_{\mathrm{map}}\) denotes the privacy mapping layer, and \(M_{\mathrm{priv}}\) denotes the protected private store. 
We use \(\Gamma\) to denote mapping-key records written to the privacy mapping layer and \(\Pi\) to denote protected records containing exact private values.

\begin{algorithm}
\caption{Privacy-Aware Memory Writing and Partitioned Storage}
\label{alg:privacy_memory_writing}
\begin{algorithmic}[1]
\Require User input \(x_t\), user ID \(u\),
sanitized vector store \(M_{\mathrm{vector}}\),
sanitized graph store \(M_{\mathrm{graph}}\),
privacy mapping layer \(M_{\mathrm{map}}\),
protected private store \(M_{\mathrm{priv}}\)
\Ensure Updated \(M_{\mathrm{vector}}\), \(M_{\mathrm{graph}}\),
\(M_{\mathrm{map}}\), and \(M_{\mathrm{priv}}\)

\State \(F_t \gets \textsc{ExtractFacts}(x_t)\)
\For{each fact object \(f=(y_f,\mathcal{A}_f) \in F_t\)}
    \If{\(\mathcal{A}_f=\emptyset\)}
        \State \(\textsc{WriteVector}(M_{\mathrm{vector}}, y_f)\)
    \Else
        \State \((\tilde{y}_f,\Gamma_f,\Pi_f)
        \gets \textsc{SanitizeFact}(f,u)\)
        \State \(\textsc{WriteVector}(M_{\mathrm{vector}}, \tilde{y}_f)\)
        \State \(\textsc{WriteMap}(M_{\mathrm{map}}, \Gamma_f)\)
        \State \(\textsc{WritePrivate}(M_{\mathrm{priv}}, \Pi_f)\)
    \EndIf
\EndFor

\State \(R_t \gets \textsc{ExtractRelations}(x_t)\)
\For{each relation triplet \(r=(s_r,\rho_r,d_r) \in R_t\)}
    \If{\(\rho_r\) does not correspond to a privacy-sensitive attribute}
        \State \(\textsc{WriteGraph}(M_{\mathrm{graph}}, r)\)
    \Else
        \State \((\tilde{r},\Gamma_r,\Pi_r)
        \gets \textsc{SanitizeRelation}(r,u)\)
        \State \(\textsc{WriteGraph}(M_{\mathrm{graph}}, \tilde{r})\)
        \State \(\textsc{WriteMap}(M_{\mathrm{map}}, \Gamma_r)\)
        \State \(\textsc{WritePrivate}(M_{\mathrm{priv}}, \Pi_r)\)
    \EndIf
\EndFor
\end{algorithmic}
\end{algorithm}

\subsection{Privacy Sanitization Strategies}
\label{app:privacy_sanitization}

After privacy-aware information extraction, detected private values are transformed into sanitized representations before being written into searchable memory, while non-private facts and triplets remain unchanged. 
To preserve task-relevant semantics without exposing exact private values, SP-Mem adopts four families of sanitization strategies: name or alias substitution, suffix-preserving masking, numerical bucketing, and LLM-based generalization. 
Each fine-grained privacy type \(\tau\) is assigned an appropriate strategy according to a type-to-strategy policy \(g\).

In the vector branch, each detected private value \(v_j\) in a fact is replaced with its sanitized value
\[
\tilde{v}_j=\textsc{Sanitize}(v_j,g(\tau_j)),
\]
producing a sanitized fact. 
In the graph branch, triplets with non-private relations remain unchanged, while for privacy-related triplets, the private destination value \(d_r\) is replaced with its sanitized counterpart
\[
\tilde{d}_r=\textsc{Sanitize}(d_r,g(\tau_r)).
\]

The four sanitization strategies are defined as follows. 
\textbf{(1) Name or alias substitution} replaces names, categorical values, or contact-related values with weaker substitutes, such as first-name fragments or synthetic aliases. 
\textbf{(2) Suffix-preserving masking} applies to structured identifiers such as phone numbers, bank accounts, credit cards, tax IDs, and passport numbers; it masks most of the value while retaining the last four digits. 
\textbf{(3) Numerical bucketing} maps numerical values into coarse semantic ranges, such as age groups, income levels, credit-score bands, or debt-ratio ranges. 
\textbf{(4) LLM-based generalization} rewrites open-ended or context-dependent private values into broader descriptions, such as generalized occupation, education, medical, relationship, transaction, insurance, or location descriptions.

Table~\ref{tab:privacy_sanitization} lists the type-to-strategy mapping used by SP-Mem.

\begin{table}[t]
\centering
\small
\setlength{\tabcolsep}{2pt}
\caption{Privacy sanitization strategies for all privacy types.}
\label{tab:privacy_sanitization}
\begin{tabularx}{\linewidth}{p{0.21\linewidth}p{0.23\linewidth}X}
\toprule
\textbf{Strategy} & \textbf{Privacy type} & \textbf{Example} \\
\midrule

\multirow{4}{0.21\linewidth}{Name/alias substitution}
& name & Maria Lombardi \(\rightarrow\) maria \\
& gender & female \(\rightarrow\) gender\_alias \\
& email & maria@example.com \(\rightarrow\) email\_alias \\
& finance status level & medium \(\rightarrow\) finance\_status\_alias \\

\midrule
\multirow{6}{0.21\linewidth}{Suffix-preserving masking}
& ID number & ID98427561 \(\rightarrow\) id\_****7561 \\
& passport number & IT-P388875140 \(\rightarrow\) passport\_******5140 \\
& phone number & +1-415-555-1289 \(\rightarrow\) phone\_******1289 \\
& bank account & 123456789012 \(\rightarrow\) bank\_****9012 \\
& credit card & 4111111111111234 \(\rightarrow\) card\_****1234 \\
& tax ID & 918273645 \(\rightarrow\) tax\_id\_*****3645 \\

\midrule
\multirow{12}{0.21\linewidth}{Numerical bucketing}
& age & 34 \(\rightarrow\) age\_adult \\
& tax payment & 6200 \(\rightarrow\) moderate\_tax\_payment \\
& monthly income & 8500 \(\rightarrow\) upper\_middle\_income \\
& monthly expenses & 3200 \(\rightarrow\) moderate\_expense \\
& account balance & 45600 \(\rightarrow\) high\_balance \\
& loan amount & 80000 \(\rightarrow\) moderate\_loan \\
& credit limit & 15000 \(\rightarrow\) high\_credit\_limit \\
& debt ratio & 0.42 \(\rightarrow\) medium\_debt\_ratio \\
& investment return & 12.4\% \(\rightarrow\) moderate\_gain \\
& ROI & 18.0\% \(\rightarrow\) high\_roi \\
& net worth & 1.2M \(\rightarrow\) high\_net\_worth \\
& credit score & 785 \(\rightarrow\) very\_good\_credit \\

\midrule
\multirow{15}{0.21\linewidth}{LLM-based generalization}
& nationality & Brazilian \(\rightarrow\) south\_america \\
& occupation & nurse \(\rightarrow\) healthcare\_professional \\
& education & master's degree \(\rightarrow\) higher\_education \\
& medical diagnosis & Primary Insomnia \(\rightarrow\) sleep\_issue \\
& medical symptoms & insomnia symptoms \(\rightarrow\) sleep\_related \\
& medical treatments & dental cleaning \(\rightarrow\) dental\_treatment \\
& medical allergy & Penicillin \(\rightarrow\) medication\_allergy \\
& medical exams & blood pressure record \(\rightarrow\) vital\_signs \\
& surgical history & appendectomy \(\rightarrow\) has\_surgery \\
& marriage & divorced \(\rightarrow\) unpartnered \\
& children count & 2 children \(\rightarrow\) has\_children \\
& transaction record & 2025-10-24 Cash - Gas Station \$45.75 \(\rightarrow\) fuel \\
& insurance record & family health coverage with liability protection \(\rightarrow\) health\_insurance \\
& home address & 1418 N Spruce Ave, Wichita, KS \(\rightarrow\) wichita\_kansas \\
& work address & 777 S Elm St, Wichita, KS \(\rightarrow\) wichita\_kansas \\

\bottomrule
\end{tabularx}
\end{table}

\subsection{Privacy-Aware Query Reasoning and Authorized Retrieval}
\label{app:inference}

Algorithm~\ref{alg:query_retrieval} summarizes the query-time
reasoning and authorized retrieval procedure of SP-Mem. Given a user query
\(q\), the query analyzer identifies the information required to complete
the task and records it as a query-specific plan \(\mathcal{P}_q\). 
It then determines whether any required entity corresponds to private
information. 
The prompt template for this query-time analysis is provided in
Figure~\ref{fig:reasoning}. 
The indicator \(b_q\) denotes whether private information is required. 
If private information is required, SP-Mem requests user consent and records the authorization decision as \(a_q\).

Retrieval is performed over the sanitized searchable memory layer:
the vector store \(M_{\mathrm{vector}}\) and the graph store
\(M_{\mathrm{graph}}\). The retrieved results are merged into
\(C_q\). If and only if the query requires private information and
user consent is granted, SP-Mem restores the task-required exact private values
through the privacy mapping layer \(M_{\mathrm{map}}\) and the protected private
store \(M_{\mathrm{priv}}\). Otherwise, the agent uses the sanitized
context directly. The final response is generated from the resulting
context \(\hat{C}_q\), using the response-generation prompt shown in
Figure~\ref{fig:response}.

\begin{algorithm}
\caption{Privacy-Aware Query Reasoning and Authorized Retrieval}
\label{alg:query_retrieval}
\begin{algorithmic}[1]
\Require Query \(q\), user ID \(u\),
sanitized vector store \(M_{\mathrm{vector}}\),
sanitized graph store \(M_{\mathrm{graph}}\),
privacy mapping layer \(M_{\mathrm{map}}\),
protected private store \(M_{\mathrm{priv}}\)
\Ensure Final response \(y\)

\State \(\mathcal{P}_q \gets \textsc{AnalyzeQuery}(q)\)
\State \(b_q \gets \textsc{NeedPrivate}(\mathcal{P}_q)\)

\If{\(b_q = 1\)}
    \State \(a_q \gets \textsc{RequestConsent}(q,u,\mathcal{P}_q)\)
\Else
    \State \(a_q \gets 0\)
\EndIf

\State \(R_v \gets \textsc{RetrieveVector}(q,\mathcal{P}_q,M_{\mathrm{vector}})\)
\State \(R_g \gets \textsc{RetrieveGraph}(q,\mathcal{P}_q,M_{\mathrm{graph}})\)
\State \(C_q \gets \textsc{Merge}(R_v,R_g)\)

\If{\(b_q = 1\) \textbf{and} \(a_q = 1\)}
    \State \(\hat{C}_q \gets
    \textsc{Restore}(C_q,M_{\mathrm{map}},M_{\mathrm{priv}})\)
\Else
    \State \(\hat{C}_q \gets C_q\)
\EndIf

\State \(y \gets \textsc{Generate}(q,\hat{C}_q)\)
\State \Return \(y\)
\end{algorithmic}
\end{algorithm}

\section{Experimental setup details}
\label{app:experimental_setup}

\paragraph{Query-time evaluation pipeline.}
Each memory-augmented method is evaluated through a query-time pipeline consisting of query analysis, consent handling, memory retrieval, and response generation. 
Given a user query, the agent identifies the task-required memory entities and determines whether exact private information is needed. 
When private information is required, the agent produces a consent request before generating the final response. 
The response is then generated from evidence retrieved from the corresponding memory backend. 
Across methods, the memory backend determines how user histories are stored and retrieved, while the query analysis, consent handling, and response-generation procedure are kept fixed.

\paragraph{Evaluation tasks.}
We evaluate five tasks: Preference-only, Privacy-only-allowed, Privacy-only-denied, Mixed-allowed, and Mixed-denied. 
Preference-only tasks require non-private user preferences only and do not require exact private information. 
Privacy-only-allowed tasks require exact private information, and the user allows privacy usage. 
Privacy-only-denied tasks require exact private information, but the user denies privacy usage. 
Mixed-allowed tasks require both non-private user preferences and exact private information, and the user allows privacy usage. 
Mixed-denied tasks require both non-private user preferences and exact private information, but the user denies privacy usage. 
P-TC is evaluated for all five tasks. 
P-PQ is evaluated for Preference-only, Mixed-allowed, and Mixed-denied tasks, where preference usage is required. 
UPU is reported for Preference-only, Privacy-only-denied, and Mixed-denied tasks, where exact private values should not appear.

\paragraph{Evaluation protocols.}
We use GPT-4.1~\citep{openai_gpt41} as the judge model for response-quality evaluation. 
For response quality, we report \textbf{pairwise task completion (P-TC)} and \textbf{pairwise personalization quality (P-PQ)}, since pairwise comparison provides a natural evaluation protocol for open-ended assistant responses and can be more robust than pointwise scoring~\citep{chiang2024chatbot,raina2024llmjudge}. 
For each evaluation query, we compare the SP-Mem response against the response from one baseline under the same task context. 
The judge is given the task description, user query, and two anonymized responses, and is asked to choose the better response according to the metric-specific rubric or return a tie. 
To reduce position bias, each comparison is evaluated twice with the response order swapped. 
We canonicalize the two judgments back to the same system identities; a non-tie winner is accepted only when the two order-swapped judgments agree, and all inconsistent cases are counted as ties. 
We then report the win-tie rate of SP-Mem, \((W+T)/N\), where \(W\), \(T\), and \(N\) denote the number of SP-Mem wins, ties, and total SP-Mem--baseline comparisons, respectively. 
For privacy behavior, \textbf{Privacy-Appropriate Requesting (PAR)} is evaluated by comparing whether the agent requests authorization against whether the task requires exact private information. 
\textbf{Unnecessary Privacy Usage (UPU)} is computed with a rule-based exact-match scorer against the user's privacy-value inventory, where masked, sanitized, or generalized values are not counted as exact exposure. 
The prompts for P-TC and P-PQ are shown in Figures~\ref{fig:tc-pairwise-prompt} and~\ref{fig:pq-pairwise-prompt}, respectively.

\paragraph{Compute resources.}
Closed-source LLM experiments, including GPT-5.2-Chat generation and GPT-4.1 judge evaluation, were conducted through API calls. 
Open-source backbone models were locally deployed and run on NVIDIA A40 GPUs. 
Memory storage and retrieval used Qdrant for vector memory and Neo4j for graph memory.

\section{Additional Experiment Details}
\label{app:additional_experiment_details}

Tables~\ref{tab:domain-education-mode-pairwise}, \ref{tab:domain-finance-mode-pairwise}, \ref{tab:domain-medical-mode-pairwise}, and~\ref{tab:domain-mental-mode-pairwise} provide pairwise results for the Education, Finance, Medical, and Mental domains, respectively. 
Each table breaks down SP-Mem's win-tie rates against each baseline by evaluation task and response backbone. 
The five evaluation tasks are Preference-only, Privacy-only-allowed, Privacy-only-denied, Mixed-allowed, and Mixed-denied.

Table~\ref{tab:full-ablation-memory} reports the full ablation results for the memory architecture. 
It compares the hybrid Vector+Graph memory against Graph-only and Vector-only variants across domains and response backbones.

\begin{table}[t]
  \caption{Full ablation results of \textbf{Vector+Graph} memory against single-branch variants.
  Each value reports the win-tie rate, $(W+T)/N$, for P-TC and P-PQ within each domain and response backbone.
  Values above 50\% indicate that \textbf{Vector+Graph} is preferred over or comparable to the corresponding single-branch variant.
  $\uparrow$ denotes higher is better.}
  \label{tab:full-ablation-memory}
  \centering
  \small
  \setlength{\tabcolsep}{2pt}
  \newlength{\ablfullcolwidth}
  \setlength{\ablfullcolwidth}{1.08cm}
  \begin{tabular*}{\linewidth}{@{\extracolsep{\fill}}
    ll
    *{4}{>{\centering\arraybackslash}p{\ablfullcolwidth} >{\centering\arraybackslash}p{\ablfullcolwidth}}
    @{}}
    \toprule
    \multirow{2}{*}{Domain}
    & \multirow{2}{*}{\textbf{Vector+Graph} vs.}
    & \multicolumn{2}{c}{GPT-5.2}
    & \multicolumn{2}{c}{Llama-3.1}
    & \multicolumn{2}{c}{DeepSeek}
    & \multicolumn{2}{c}{Qwen3} \\
    \cmidrule(lr){3-4}
    \cmidrule(lr){5-6}
    \cmidrule(lr){7-8}
    \cmidrule(lr){9-10}
    & & P-TC$\uparrow$ & P-PQ$\uparrow$
      & P-TC$\uparrow$ & P-PQ$\uparrow$
      & P-TC$\uparrow$ & P-PQ$\uparrow$
      & P-TC$\uparrow$ & P-PQ$\uparrow$ \\
    \midrule
    \multirow{2}{*}{Education}
    & Graph-only  & 90.99\% & 86.87\% & 85.80\% & 81.04\% & 88.89\% & 85.22\% & 86.79\% & 80.60\% \\
    & Vector-only & 87.52\% & 91.03\% & 85.19\% & 86.87\% & 88.15\% & 90.30\% & 87.56\% & 89.01\% \\
    \midrule
    \multirow{2}{*}{Finance}
    & Graph-only  & 85.80\% & 85.71\% & 84.20\% & 78.25\% & 89.63\% & 86.19\% & 88.15\% & 83.33\% \\
    & Vector-only & 89.63\% & 90.63\% & 84.14\% & 86.46\% & 87.78\% & 91.59\% & 85.68\% & 87.46\% \\
    \midrule
    \multirow{2}{*}{Medical}
    & Graph-only  & 95.68\% & 93.49\% & 87.16\% & 81.11\% & 88.15\% & 85.24\% & 85.93\% & 83.02\% \\
    & Vector-only & 94.07\% & 93.65\% & 87.27\% & 85.02\% & 89.26\% & 90.32\% & 87.41\% & 86.83\% \\
    \midrule
    \multirow{2}{*}{Mental}
    & Graph-only  & 91.98\% & 88.96\% & 86.26\% & 80.99\% & 87.53\% & 84.93\% & 84.69\% & 82.24\% \\
    & Vector-only & 91.85\% & 94.33\% & 85.52\% & 81.59\% & 88.52\% & 90.45\% & 88.27\% & 89.55\% \\
    \bottomrule
  \end{tabular*}
\end{table}

\begin{table}[p]
  \caption{Pairwise results on the Education domain by evaluation task.
Each cell reports the P-TC/P-PQ win-tie rate of \textbf{SP-Mem} against the corresponding baseline under the same response backbone.
For Privacy-only-allowed and Privacy-only-denied tasks, P-PQ is not applicable and is shown as --.
Values above 50\% indicate that \textbf{SP-Mem} is preferred over or comparable to the baseline.}
  \label{tab:domain-education-mode-pairwise}
  \centering
  \small
  \setlength{\tabcolsep}{2pt}
  \begin{tabular*}{\linewidth}{@{\extracolsep{\fill}} l l c c c c @{}}
    \toprule
    \multicolumn{6}{c}{\textbf{Education}} \\
    \midrule
    \multirow{2}{*}{Task} & \multirow{2}{*}{\textbf{SP-Mem} vs.}
    & GPT-5.2 & Llama-3.1 & DeepSeek & Qwen3 \\
    & & P-TC/P-PQ & P-TC/P-PQ & P-TC/P-PQ & P-TC/P-PQ \\
    \midrule
    Preference-only & Full-context & 72.44\%/84.30\% & 67.26\%/85.63\% & 81.93\%/84.74\% & 47.70\%/70.52\% \\
     & MemOS & 87.56\%/95.56\% & 84.59\%/91.26\% & 93.48\%/95.11\% & 68.30\%/90.07\% \\
     & Mem0 & 89.19\%/93.19\% & 81.48\%/83.11\% & 95.26\%/95.70\% & 72.44\%/87.85\% \\
     & Zep & 89.90\%/94.95\% & 83.11\%/89.93\% & 94.79\%/97.32\% & 70.07\%/90.52\% \\
    \midrule
    Privacy-only-allowed & Full-context & 61.71\%/-- & 76.57\%/-- & 64.00\%/-- & 44.57\%/-- \\
     & MemOS & 86.29\%/-- & 89.14\%/-- & 94.86\%/-- & 79.43\%/-- \\
     & Mem0 & 90.86\%/-- & 93.14\%/-- & 93.71\%/-- & 81.71\%/-- \\
     & Zep & 94.80\%/-- & 92.57\%/-- & 99.43\%/-- & 89.14\%/-- \\
    \midrule
    Privacy-only-denied & Full-context & 98.29\%/-- & 96.57\%/-- & 93.71\%/-- & 92.57\%/-- \\
     & MemOS & 96.57\%/-- & 92.00\%/-- & 96.00\%/-- & 92.00\%/-- \\
     & Mem0 & 93.14\%/-- & 93.14\%/-- & 97.14\%/-- & 96.00\%/-- \\
     & Zep & 95.95\%/-- & 93.71\%/-- & 98.29\%/-- & 96.00\%/-- \\
    \midrule
    Mixed-allowed & Full-context & 50.00\%/69.20\% & 70.00\%/83.80\% & 60.20\%/73.20\% & 40.80\%/69.60\% \\
     & MemOS & 92.00\%/94.20\% & 88.60\%/87.40\% & 97.20\%/97.00\% & 81.20\%/87.80\% \\
     & Mem0 & 93.80\%/93.20\% & 88.60\%/83.00\% & 98.40\%/95.80\% & 84.80\%/87.80\% \\
     & Zep & 93.56\%/97.18\% & 93.80\%/87.60\% & 97.60\%/95.60\% & 85.40\%/89.20\% \\
    \midrule
    Mixed-denied & Full-context & 88.20\%/88.60\% & 96.40\%/98.60\% & 85.40\%/90.20\% & 87.80\%/89.40\% \\
     & MemOS & 91.80\%/94.00\% & 96.80\%/92.60\% & 95.60\%/95.60\% & 85.40\%/92.60\% \\
     & Mem0 & 91.80\%/94.20\% & 94.80\%/88.00\% & 94.20\%/93.60\% & 89.80\%/90.40\% \\
     & Zep & 92.56\%/94.77\% & 98.20\%/93.20\% & 95.00\%/95.20\% & 88.60\%/93.60\% \\
    \bottomrule
  \end{tabular*}
\end{table}

\begin{table}[p]
  \caption{Pairwise results on the Finance domain by evaluation task.
  Each cell reports the P-TC/P-PQ win-tie rate of \textbf{SP-Mem} against the corresponding baseline under the same response backbone.
  For Privacy-only-allowed and Privacy-only-denied tasks, P-PQ is not applicable and is shown as --.
  Values above 50\% indicate that \textbf{SP-Mem} is preferred over or comparable to the baseline.}
  \label{tab:domain-finance-mode-pairwise}
  \centering
  \small
  \setlength{\tabcolsep}{2pt}
  \begin{tabular*}{\linewidth}{@{\extracolsep{\fill}} l l c c c c @{}}
    \toprule
    \multicolumn{6}{c}{\textbf{Finance}} \\
    \midrule
    \multirow{2}{*}{Task} & \multirow{2}{*}{\textbf{SP-Mem} vs.}
    & GPT-5.2 & Llama-3.1 & DeepSeek & Qwen3 \\
    & & P-TC/P-PQ & P-TC/P-PQ & P-TC/P-PQ & P-TC/P-PQ \\
    \midrule
    Preference-only & Full-context & 65.23\%/79.20\% & 58.67\%/83.41\% & 79.97\%/81.60\% & 48.35\%/73.87\% \\
     & MemOS & 88.30\%/92.74\% & 80.15\%/89.48\% & 89.04\%/92.44\% & 69.04\%/89.63\% \\
     & Mem0 & 87.70\%/93.04\% & 84.30\%/88.89\% & 92.59\%/95.11\% & 77.93\%/92.44\% \\
     & Zep & 88.07\%/92.60\% & 81.78\%/88.89\% & 91.96\%/94.23\% & 69.15\%/89.82\% \\
    \midrule
    Privacy-only-allowed & Full-context & 55.11\%/-- & 61.78\%/-- & 59.38\%/-- & 43.56\%/-- \\
     & MemOS & 79.11\%/-- & 80.89\%/-- & 91.56\%/-- & 80.44\%/-- \\
     & Mem0 & 86.22\%/-- & 89.78\%/-- & 95.56\%/-- & 85.33\%/-- \\
     & Zep & 91.20\%/-- & 89.78\%/-- & 95.43\%/-- & 86.51\%/-- \\
    \midrule
    Privacy-only-denied & Full-context & 97.78\%/-- & 96.44\%/-- & 95.54\%/-- & 91.11\%/-- \\
     & MemOS & 92.44\%/-- & 90.22\%/-- & 94.67\%/-- & 95.11\%/-- \\
     & Mem0 & 95.11\%/-- & 93.33\%/-- & 96.89\%/-- & 93.33\%/-- \\
     & Zep & 96.30\%/-- & 95.56\%/-- & 96.80\%/-- & 93.95\%/-- \\
    \midrule
    Mixed-allowed & Full-context & 43.33\%/70.44\% & 56.00\%/82.89\% & 59.11\%/74.22\% & 40.76\%/72.61\% \\
     & MemOS & 84.67\%/90.67\% & 83.07\%/88.86\% & 92.67\%/92.89\% & 74.39\%/88.86\% \\
     & Mem0 & 89.56\%/92.44\% & 89.78\%/86.22\% & 96.00\%/93.11\% & 84.89\%/90.00\% \\
     & Zep & 89.30\%/89.77\% & 89.78\%/87.11\% & 96.15\%/92.29\% & 83.86\%/88.18\% \\
    \midrule
    Mixed-denied & Full-context & 85.33\%/89.33\% & 95.11\%/94.89\% & 85.33\%/84.22\% & 84.44\%/91.56\% \\
     & MemOS & 87.56\%/91.33\% & 95.55\%/93.54\% & 91.11\%/90.44\% & 83.52\%/92.43\% \\
     & Mem0 & 88.00\%/91.33\% & 92.67\%/89.11\% & 95.78\%/91.56\% & 90.00\%/91.56\% \\
     & Zep & 90.00\%/90.47\% & 96.00\%/90.89\% & 94.33\%/89.57\% & 86.14\%/90.68\% \\
    \bottomrule
  \end{tabular*}
\end{table}

\begin{table}[p]
  \caption{Pairwise results on the Medical domain by evaluation task.
  Each cell reports the P-TC/P-PQ win-tie rate of \textbf{SP-Mem} against the corresponding baseline under the same response backbone.
  For Privacy-only-allowed and Privacy-only-denied tasks, P-PQ is not applicable and is shown as --.
  Values above 50\% indicate that \textbf{SP-Mem} is preferred over or comparable to the baseline.}
  \label{tab:domain-medical-mode-pairwise}
  \centering
  \small
  \setlength{\tabcolsep}{2pt}
  \begin{tabular*}{\linewidth}{@{\extracolsep{\fill}} l l c c c c @{}}
    \toprule
    \multicolumn{6}{c}{\textbf{Medical}} \\
    \midrule
    \multirow{2}{*}{Task} & \multirow{2}{*}{\textbf{SP-Mem} vs.}
    & GPT-5.2 & Llama-3.1 & DeepSeek & Qwen3 \\
    & & P-TC/P-PQ & P-TC/P-PQ & P-TC/P-PQ & P-TC/P-PQ \\
    \midrule
    Preference-only & Full-context & 83.66\%/87.29\% & 69.33\%/89.93\% & 82.47\%/89.00\% & 52.19\%/77.07\% \\
     & MemOS & 92.30\%/97.04\% & 88.69\%/96.28\% & 94.37\%/97.63\% & 77.63\%/95.70\% \\
     & Mem0 & 91.41\%/95.11\% & 86.22\%/89.93\% & 94.96\%/97.93\% & 77.19\%/91.11\% \\
     & Zep & 91.56\%/95.85\% & 82.96\%/88.44\% & 93.40\%/96.25\% & 73.13\%/91.79\% \\
    \midrule
    Privacy-only-allowed & Full-context & 57.92\%/-- & 66.67\%/-- & 61.16\%/-- & 33.63\%/-- \\
     & MemOS & 92.89\%/-- & 88.89\%/-- & 98.22\%/-- & 89.78\%/-- \\
     & Mem0 & 89.33\%/-- & 86.22\%/-- & 95.11\%/-- & 83.56\%/-- \\
     & Zep & 96.10\%/-- & 92.00\%/-- & 99.08\%/-- & 90.58\%/-- \\
    \midrule
    Privacy-only-denied & Full-context & 94.59\%/-- & 97.78\%/-- & 86.10\%/-- & 90.54\%/-- \\
     & MemOS & 95.11\%/-- & 92.89\%/-- & 96.44\%/-- & 91.11\%/-- \\
     & Mem0 & 91.11\%/-- & 91.11\%/-- & 92.44\%/-- & 90.22\%/-- \\
     & Zep & 98.05\%/-- & 96.89\%/-- & 95.87\%/-- & 90.13\%/-- \\
    \midrule
    Mixed-allowed & Full-context & 67.26\%/80.04\% & 71.11\%/88.89\% & 63.98\%/77.85\% & 43.21\%/73.94\% \\
     & MemOS & 99.78\%/99.11\% & 95.96\%/94.84\% & 99.11\%/99.56\% & 92.44\%/95.78\% \\
     & Mem0 & 97.11\%/95.56\% & 91.11\%/83.78\% & 97.56\%/96.44\% & 85.56\%/90.22\% \\
     & Zep & 97.48\%/97.12\% & 94.22\%/89.56\% & 99.10\%/96.85\% & 88.62\%/94.20\% \\
    \midrule
    Mixed-denied & Full-context & 94.12\%/95.02\% & 96.00\%/96.22\% & 87.75\%/89.53\% & 83.26\%/91.29\% \\
     & MemOS & 96.21\%/98.44\% & 96.86\%/95.74\% & 96.00\%/97.11\% & 86.67\%/96.44\% \\
     & Mem0 & 95.56\%/96.44\% & 93.78\%/88.89\% & 96.22\%/93.78\% & 88.67\%/91.11\% \\
     & Zep & 95.68\%/97.12\% & 95.78\%/93.33\% & 95.50\%/95.72\% & 86.61\%/94.20\% \\
    \bottomrule
  \end{tabular*}
\end{table}

\begin{table}[p]
  \caption{Pairwise results on the Mental domain by evaluation task.
  Each cell reports the P-TC/P-PQ win-tie rate of \textbf{SP-Mem} against the corresponding baseline under the same response backbone.
  For Privacy-only-allowed and Privacy-only-denied tasks, P-PQ is not applicable and is shown as --.
  Values above 50\% indicate that \textbf{SP-Mem} is preferred over or comparable to the baseline.}
  \label{tab:domain-mental-mode-pairwise}
  \centering
  \small
  \setlength{\tabcolsep}{2pt}
  \begin{tabular*}{\linewidth}{@{\extracolsep{\fill}} l l c c c c @{}}
    \toprule
    \multicolumn{6}{c}{\textbf{Mental}} \\
    \midrule
    \multirow{2}{*}{Task} & \multirow{2}{*}{\textbf{SP-Mem} vs.}
    & GPT-5.2 & Llama-3.1 & DeepSeek & Qwen3 \\
    & & P-TC/P-PQ & P-TC/P-PQ & P-TC/P-PQ & P-TC/P-PQ \\
    \midrule
    Preference-only & Full-context & 79.73\%/80.95\% & 66.51\%/84.63\% & 81.14\%/81.44\% & 48.11\%/68.23\% \\
     & MemOS & 92.30\%/95.56\% & 82.96\%/88.59\% & 93.04\%/95.41\% & 73.48\%/88.89\% \\
     & Mem0 & 91.11\%/95.56\% & 83.85\%/86.81\% & 93.48\%/96.00\% & 76.30\%/89.63\% \\
     & Zep & 90.22\%/94.67\% & 78.81\%/88.74\% & 93.77\%/95.99\% & 70.96\%/89.19\% \\
    \midrule
    Privacy-only-allowed & Full-context & 64.74\%/-- & 72.00\%/-- & 66.47\%/-- & 39.18\%/-- \\
     & MemOS & 82.29\%/-- & 82.86\%/-- & 94.86\%/-- & 82.29\%/-- \\
     & Mem0 & 85.14\%/-- & 87.43\%/-- & 94.29\%/-- & 83.43\%/-- \\
     & Zep & 95.68\%/-- & 93.14\%/-- & 100.00\%/-- & 93.71\%/-- \\
    \midrule
    Privacy-only-denied & Full-context & 98.86\%/-- & 94.29\%/-- & 91.67\%/-- & 93.60\%/-- \\
     & MemOS & 94.86\%/-- & 89.14\%/-- & 98.29\%/-- & 97.14\%/-- \\
     & Mem0 & 96.00\%/-- & 90.86\%/-- & 97.71\%/-- & 94.29\%/-- \\
     & Zep & 95.68\%/-- & 97.14\%/-- & 99.43\%/-- & 93.14\%/-- \\
    \midrule
    Mixed-allowed & Full-context & 55.56\%/75.76\% & 69.92\%/86.59\% & 60.00\%/72.24\% & 41.53\%/66.94\% \\
     & MemOS & 94.20\%/94.60\% & 87.20\%/82.72\% & 94.20\%/96.00\% & 81.80\%/83.20\% \\
     & Mem0 & 93.80\%/95.80\% & 94.11\%/83.13\% & 97.40\%/95.40\% & 88.60\%/86.40\% \\
     & Zep & 95.02\%/95.95\% & 92.68\%/82.72\% & 96.60\%/95.60\% & 87.78\%/83.57\% \\
    \midrule
    Mixed-denied & Full-context & 88.57\%/90.82\% & 96.14\%/97.15\% & 82.79\%/86.23\% & 82.24\%/83.67\% \\
     & MemOS & 92.80\%/94.60\% & 93.09\%/84.35\% & 95.20\%/93.20\% & 84.40\%/86.40\% \\
     & Mem0 & 93.80\%/96.00\% & 96.14\%/86.18\% & 95.80\%/95.20\% & 89.00\%/89.00\% \\
     & Zep & 90.65\%/96.26\% & 95.12\%/88.82\% & 95.00\%/93.60\% & 87.78\%/84.37\% \\
    \bottomrule
  \end{tabular*}
\end{table}

\begin{figure}[t]
\centering
\begin{tcolorbox}[
    colback=white,
    colframe=black,
    boxrule=0.6pt,
    arc=0pt,
    width=\linewidth,
    left=6mm,
    right=6mm,
    top=3mm,
    bottom=3mm
]
\small

You are a Preference Selector. Your task is to analyze a user's demographic and lifestyle profile, then select the most appropriate preferences from a given bank of non-private preference candidates.

\medskip
\textbf{Task}

Given a user profile and a preference candidate bank across multiple preference dimensions, analyze the user's characteristics and select the top-\(N\) best-matching preference values for each dimension.

\medskip
\textbf{Profile Information to Consider}

When selecting preferences, consider the following information from the user profile:

\begin{itemize}[leftmargin=1.5em, itemsep=0.15em, topsep=0.2em]
    \item \textbf{Demographics:} age, gender, nationality, occupation, and education.
    \item \textbf{Medical history:} past diagnosis, symptoms, and allergies. Medical history should be used only for preference refinement, not as a hard restriction.
    \item \textbf{Lifestyle:} marriage status, children count, monthly income, and finance status level.
\end{itemize}

\medskip
\textbf{Selection Requirements}

\begin{itemize}[leftmargin=1.5em, itemsep=0.15em, topsep=0.2em]
    \item Select preferences that are realistic, personalized, and diverse.
    \item Avoid stereotypes or overly deterministic assumptions from demographic attributes.
    \item Strictly select values from the provided candidate lists.
    \item Each selected value must be copied verbatim from the corresponding candidate list.
    \item Do not invent new preferences or include private identifiers.
    \item Output valid JSON only, with exactly one key for each preference dimension.
\end{itemize}

\medskip
\textbf{Input}

User Privacy Profile:

[user\_privacy\_profile]

\medskip
Preference Candidate Bank:

[candidates\_by\_dimension]

\medskip

\textbf{Output Format}

\begin{verbatim}
{
  "shows": ["<preference 1>", "<preference 2>", "..."],
  "music": ["<preference 1>", "<preference 2>", "..."]
}
\end{verbatim}

\medskip
The system then selects the least-used item from each ranked list to improve preference diversity across users.

\end{tcolorbox}
\caption{The prompt used for privacy-conditioned preference selection. Given a user profile and a candidate bank, the model ranks non-private preference values for each dimension and returns structured JSON for controlled parsing.}
\label{fig:preference_selection_prompt}
\end{figure}

\begin{figure}[t]
\centering
\begin{tcolorbox}[
    colback=white,
    colframe=black,
    boxrule=0.6pt,
    arc=0pt,
    width=\linewidth,
    left=4mm,
    right=4mm,
    top=3mm,
    bottom=3mm
]
\small
You are a Privacy Profile Consistency Checker. Your task is to examine a single user privacy profile and identify only hard logical contradictions.

\medskip
\textbf{Core Rule}

Flag an issue only when two or more fields are logically incompatible or explicitly self-contradictory. Do not flag a profile because a value is uncommon, atypical, or statistically unlikely.

\medskip
\textbf{Checks}

\begin{itemize}[leftmargin=1.4em, itemsep=0.12em, topsep=0.15em]
    \item \textbf{Age--education--occupation:} detect impossible combinations, such as a child with a PhD or a senior professional role.
    \item \textbf{Numeric and financial fields:} detect invalid ratios, negative values for fields that cannot be negative, or mutually impossible numeric constraints.
    \item \textbf{Medical sex/gender consistency:} detect strict organ-specific contradictions, such as pregnancy-related conditions in a male profile or prostate/testicle-related conditions in a female profile.
    \item \textbf{Relationship fields:} detect explicit self-contradictions, such as a single marital status together with a spouse field.
    \item \textbf{Demographic fields:} detect direct conflicts between explicit demographic attributes, such as inconsistent nationality or citizenship fields.
\end{itemize}

\medskip
\textbf{Do Not Flag}

Do not infer contradictions from names, addresses, language, insurance records, or treatments that are merely unusual. Do not make judgments based on stereotypes or typical life patterns.

\medskip
\textbf{Input}

User Privacy Profile:

[user\_privacy\_profile]

\medskip
\textbf{Output Format}

{\ttfamily\small
\begin{tabular}{@{}l@{}}
\{ \\
\hspace*{1.5em}"profile\_conflict\_label": 0 or 1, \\
\hspace*{1.5em}"issues": ["category: description", ...] \\
\}
\end{tabular}
}

\medskip
Set \textit{profile\_conflict\_label} to 1 if and only if the issue list is non-empty. If no hard contradiction is found, return an empty issue list.

\end{tcolorbox}
\caption{The prompt for privacy-profile consistency checking. The checker identifies only hard logical contradictions within a user privacy profile and returns structured JSON for controlled parsing.}
\label{fig:privacy_consistency_prompt}
\end{figure}

\begin{figure}[t]
\centering
\begin{tcolorbox}[
    colback=white,
    colframe=black,
    boxrule=0.6pt,
    arc=0pt,
    width=\linewidth,
    left=4mm,
    right=4mm,
    top=3mm,
    bottom=3mm
]
\small

You are a Privacy--Preference Consistency Checker. Your task is to examine a synthetic user persona and determine whether the user's privacy profile and non-private preferences contain any direct contradiction.

\medskip
\textbf{Goal}

Given two views of the same user, decide whether the assigned preferences are consistent with the user's background facts. A conflict should be reported only when the two views cannot logically hold at the same time.

\medskip
\textbf{Conflict Definition}

Label a conflict only if the information is mutually exclusive, logically incompatible, or factually contradictory. Do not label a profile as inconsistent merely because a preference is unusual, low-probability, or statistically atypical.

\medskip
\textbf{Checks}

\begin{itemize}[leftmargin=1.4em, itemsep=0.12em, topsep=0.15em]
    \item \textbf{Within-domain consistency:} check whether privacy facts and preferences in the same domain directly contradict each other.
    \item \textbf{Cross-domain consistency:} check whether facts from one domain make preferences from another domain logically impossible.
    \item \textbf{Medical interpretation:} treat medical information as medical history, not necessarily as current functional limitation.
    \item \textbf{Direct incompatibility:} flag only conflicts that are explicitly supported by the provided fields.
\end{itemize}

\medskip
\textbf{Examples}

\begin{itemize}[leftmargin=1.4em, itemsep=0.12em, topsep=0.15em]
    \item A five-year-old user with an occupation as a doctor is a conflict.
    \item A user who is explicitly unable to walk but prefers running is a conflict.
    \item A user with a history of knee surgery who prefers running is not necessarily a conflict.
\end{itemize}

\medskip
\textbf{Input}

Privacy Profile:

[privacy\_profile]

\medskip
Non-private Preferences:

[non\_privacy\_preferences]

\medskip
\textbf{Output Format}

\begin{Verbatim}[fontsize=\small]
{
  "label": 0 or 1,
  "conflicts": [
    {
      "type": "concise conflict tag",
      "reason": "brief explanation of direct incompatibility"
    }
  ],
  "explanation": "short summary of why label is 0 or 1"
}
\end{Verbatim}

\medskip
Set \textit{label} to 1 only if at least one direct conflict exists. If no conflict is found, set \textit{label} to 0 and return an empty conflict list.

\end{tcolorbox}
\caption{The prompt used for privacy-preference consistency checking. The checker identifies direct contradictions between privacy attributes and non-private preferences and returns structured JSON for controlled parsing.}
\label{fig:privacy_preference_consistency_prompt}
\end{figure}

\begin{figure}[t]
\centering
\begin{tcolorbox}[
    colback=white,
    colframe=black,
    boxrule=0.6pt,
    arc=0pt,
    width=\linewidth,
    left=4mm,
    right=4mm,
    top=3mm,
    bottom=3mm
]
\small

You are a Task-Card Generator. Your task is to generate realistic user--assistant task cards that require personal information, user preferences, or both to deliver personalized assistance.

\medskip
\textbf{Input}

\begin{itemize}[leftmargin=1.4em, itemsep=0.12em, topsep=0.15em]
    \item Scenario: the scenario name and its description.
    \item Mode: one of \textit{privacy\_only}, \textit{non\_privacy\_only}, or \textit{mixed}.
    \item Fixed privacy entities: the privacy entities that must be used in the task.
    \item Fixed preference entities: the preference entities that must be used in the task.
\end{itemize}

\medskip
Example

[Example]

\medskip
\textbf{Requirements}

\begin{enumerate}[leftmargin=1.5em, itemsep=0.12em, topsep=0.15em]
    \item Broad and reusable: describe a general task that can apply to different users and must not contain concrete personal details.
    \item Goal-driven: the task should have a clear outcome that indicates successful completion.
    \item Entity-dependent: every fixed privacy or preference entity must be necessary for completing the task.
    \item Generic only: do not instantiate concrete privacy values, identifiers, locations, medical values, financial values, or concrete preference values.
    \item Authentic and real-world: the task should reflect a common situation where users would genuinely ask an assistant for help.
    \item First-party privacy: all privacy entities must refer to the user, not to another person.
    \item Diversity: vary scenarios, goals, and workflows to avoid repetitive task patterns.
\end{enumerate}

\medskip
\textbf{Output}

Return exactly one JSON object with the following fields:

\begin{itemize}[leftmargin=1.4em, itemsep=0.12em, topsep=0.15em]
    \item scenario: the fixed scenario identifier.
    \item task\_description: a broad, reusable task description that explains why the selected entities are needed.
    \item user\_intent: a natural user request that initiates the conversation.
    \item privacy\_entities: the fixed privacy entity list.
    \item preference\_entities: the fixed preference entity list.
    \item mode: the fixed task mode.
\end{itemize}

\end{tcolorbox}
\caption{The prompt used for constrained task-card generation. The generator uses a scenario, a mode, fixed entity annotations, and a few-shot example to produce realistic task cards while preserving ground-truth privacy and preference labels.}
\label{fig:task_card_generation_prompt}
\end{figure}

\begin{figure}[t]
\centering
\begin{tcolorbox}[
    colback=white,
    colframe=black,
    boxrule=0.6pt,
    arc=0pt,
    width=\linewidth,
    left=4mm,
    right=4mm,
    top=3mm,
    bottom=3mm
]
\small

You are an expert memory-planning analyst for a privacy-aware personal assistant. 
You are precise, conservative with private fields, and deterministic.

\medskip
\textbf{Task}

Convert a user query into a minimal memory-retrieval plan. 
Decide what user-specific entities are truly needed to answer the query.

\medskip
\textbf{Allowed Entity Set}

[allowed\_entities]

\medskip
\textbf{Privacy Entity Set}

[privacy\_entities]

\medskip
\textbf{Reasoning Framework}

Follow the questions below in order and reflect them in \textit{reasoning\_process}:

\begin{description}[leftmargin=1.8em, itemsep=0.12em, topsep=0.15em]
    \item[\textbf{Q1. What is the user task?}] State the concrete task the user wants to complete.
    \item[\textbf{Q2. What information is needed?}] List only task-critical user information, map it to the allowed entity vocabulary, and remove nice-to-have extras.
    \item[\textbf{Q3. Which information is private?}] Mark an entity as private only if it belongs to the privacy entity set, and set \textit{needs\_privacy} to true iff any selected entity is private.
    \item[\textbf{Q4. Safety check.}] Do not infer extra private entities or hidden sensitive values that are not required by the task.
\end{description}

\medskip
\textbf{Hard Constraints}

All selected entities must come from the whitelist, and no extra private entities should be inferred unless they are explicitly required by the user query. 
The \textit{reasoning\_process} must contain 4--6 ordered steps covering Q1--Q3 and the final \textit{needs\_privacy} decision. 
The model must return valid JSON only, without markdown, prose, or code fences.

\medskip
\textbf{Few-shot Examples}

[Example 1]

[Example 2]

\medskip
\textbf{User Query}

[query]

\medskip
\textbf{Output JSON Schema}

\begin{verbatim}
{
  "intent": "short_intent_name",
  "reasoning_process": ["step 1", "step 2", "step 3"],
  "reasoning_summary": "brief rationale",
  "required_entities": [
    {
      "entity": "loan_amount",
      "privacy": true
    }
  ],
  "needs_privacy": true
}
\end{verbatim}

\end{tcolorbox}
\caption{The prompt used for query-time analysis.}
\label{fig:reasoning}
\end{figure}

\begin{figure}[t]
\centering
\begin{tcolorbox}[
    colback=white,
    colframe=black,
    boxrule=0.6pt,
    arc=0pt,
    width=\linewidth,
    left=4mm,
    right=4mm,
    top=3mm,
    bottom=3mm
]
\small

You are a helpful, thoughtful personal assistant. 
Your job is to write natural, high-quality answers grounded in retrieved memory evidence.

\medskip
\textbf{Primary Objective}

Answer the user query completely, accurately, and in a user-friendly style. 
When writing the answer, combine the user query with the required entities and retrieved memories.

\medskip
\textbf{User Query}

[query]

\medskip
\textbf{Retrieved Memories}

[retrieved memories]

\medskip
\textbf{Required Entities}

[required entities]

\medskip
\textbf{Execution Rules}

\begin{enumerate}[leftmargin=1.5em, itemsep=0.12em, topsep=0.15em]
    \item \textbf{Evidence grounding:} use only facts present in retrieved memories and never invent user details.
    \item \textbf{Coverage:} inspect every retrieved entity group and use all relevant evidence. For each required entity, integrate it if evidence exists; if evidence is missing and matters, explicitly state the gap.
    \item \textbf{Preference alignment:} if retrieved evidence includes user preferences, adapt wording, tone, examples, and recommendations to those preferences. Preference alignment must be substantive, not merely a mention of preferences.
    \item \textbf{Output format:} output the final answer text only.
\end{enumerate}

\medskip
Generate the final answer from the structured input. 
Use the query, required entities, and retrieved memories together when composing the answer. 
Personalize the answer using retrieved user preferences in a concrete, natural way.

\medskip
\textbf{Output}

[agent reponse]

\end{tcolorbox}
\caption{The prompt used for agent response generation.}
\label{fig:response}
\end{figure}

\begin{figure}[t]
\centering
\fbox{
\begin{minipage}{\linewidth}
\small

\medskip
You are an impartial judge for pairwise evaluation of conversational assistants.

Compare Assistant A and Assistant B for Task Completion (TC).

\medskip
\textbf{Judging principles:}
\begin{itemize}
  \item Evaluate whether the response completes the user task and required deliverables.
  \item Focus on instruction-following, completeness, relevance, and practical usefulness.
  \item Do not prefer an answer because it is longer, more formal, or more verbose.
  \item Do not let position (A vs B) influence your decision.
  \item Output winner = ``A'' if Assistant A is clearly better on TC.
  \item Output winner = ``B'' if Assistant B is clearly better on TC.
  \item If both are similarly good or similarly weak, output TIE.
\end{itemize}

\medskip
\textbf{Task Description}

\{scenario\}

\medskip
\textbf{User Query}

\{query\}

\medskip
\textbf{Required Entities (for completion)}

\{required\_entities\}

\medskip
\textbf{Assistant A Response}

\{response\_a\}

\medskip
\textbf{Assistant B Response}

\{response\_b\}

\medskip
Return ONLY JSON:

\texttt{\{"winner":"A|B|TIE","reason":"one concise sentence"\}}
\end{minipage}
}
\caption{Pairwise evaluation prompt for Task Completion (P-TC). }

\label{fig:tc-pairwise-prompt}
\end{figure}

\begin{figure}[t]
\centering
\fbox{
\begin{minipage}{\linewidth}
\small

\medskip
You are an impartial judge for pairwise evaluation of conversational assistants.

Compare Assistant A and Assistant B on Personalization Quality (PQ).

\medskip
\textbf{Judging principles:}
\begin{itemize}
  \item Evaluate whether the response uses user preferences correctly and meaningfully.
  \item ``Meaningful'' means preferences materially shape content, tone, examples, or recommendations.
  \item Penalize hallucinated, incorrect, or irrelevant preference usage.
  \item Do not prefer an answer because it is longer, more formal, or more verbose.
  \item Do not let response position (A vs B) influence your decision.
  \item Output winner = ``A'' if Assistant A is clearly better on PQ.
  \item Output winner = ``B'' if Assistant B is clearly better on PQ.
  \item Output winner = ``TIE'' if both are similarly good or similarly weak.
\end{itemize}

\medskip
\textbf{Required Preference Entities}

\{preference\_entities\}

\medskip
\textbf{User Preference Values (Ground Truth)}

\{preference\_values\}

\medskip
\textbf{User Query}

\{query\}

\medskip
\textbf{Assistant A Response}

\{response\_a\}

\medskip
\textbf{Assistant B Response}

\{response\_b\}

\medskip
Return ONLY JSON:

\texttt{\{"winner":"A|B|TIE","reason":"one concise sentence"\}}
\end{minipage}
}
\caption{Pairwise evaluation prompt for Personalization Quality (P-PQ).}
\label{fig:pq-pairwise-prompt}
\end{figure}

\end{document}